\documentclass[preprint]{aastex}

\received{}
\accepted{}
\journalid{}{}
\articleid{}{}

\shortauthors{Carpenter}
\shorttitle{2MASS Observations of Molecular Clouds}

\newcommand{\ts}{\thinspace}
\newcommand{\simless}{\mathbin{\lower 3pt\hbox
     {$\rlap{\raise 5pt\hbox{$\char'074$}}\mathchar"7218$}}}
\newcommand{\simgreat}{\mathbin{\lower 3pt\hbox
     {$\rlap{\raise 5pt\hbox{$\char'076$}}\mathchar"7218$}}}

\newcommand{\about}    {$\approx$\ts}
\newcommand{\aboutless}{$\simless$\ts}
\newcommand{\aboutmore}{$\simgreat$\ts}

\newcommand{\JB}{$J$}
\newcommand{\HB}{$H$}
\newcommand{\KB}{$K_s$}
\newcommand{\JK}{$J-K_s$}
\newcommand{\KBAND}{$K$}
\newcommand{\MASS}{{\it 2MASS}}
\newcommand{\IRAS}{{\it IRAS}}

\newcommand{\M}{$^{\rm m}$}

\newcommand{\msun}{\ts M$_\odot$}
\newcommand{\lsun}{\ts L$_\odot$}
\newcommand{\sqamin}{\ts arcmin$^{-2}$}
\newcommand{\cmmag}{cm$^{-2}$~mag$^{-1}$}

\newcommand{\etal}{et~al.}

\newcommand{\co}{$^{12}$CO}
\newcommand{\thco}{$^{13}$CO}
\newcommand{\thcoj}{$^{13}$CO(1--0)}

\newcommand{\kms}{\ts km\ts s$^{-1}$}

\newcommand{\pc}{\ts pc}

\newcommand{\hii}{H~{\sc ii}}

\newcommand{\NGC}{NGC\ts}
\newcommand{\LDN}{L\ts}
\newcommand{\GGD}{GGD\ts}
\newcommand{\IC}{IC\ts}

\def\insertplot#1#2#3#4#5#6#7{
\vskip 10pt\nobreak\hbox to \hsize{\hss\dimen0=#3in\hbox to #6\dimen0{%
\dimen0=#2in\vbox to #6\dimen0{\vss
\special{ps: plotfile #1}
\special{ps::[end]
  PGPLOT restore
}
}\hss}\hss}\vskip 10pt}

\begin{document}

\title{\MASS\ Observations of the Perseus, Orion A, Orion B, and 
       Monoceros~R2 Molecular Clouds}

\author{John M. Carpenter\\
        California Institute of Technology, 
        Department of Astronomy, MS 105-24, \\ Pasadena, CA 91125; 
        email: jmc@astro.caltech.edu}

\begin{abstract}

We use the \MASS\ Second Incremental Release Point Source Catalog to 
investigate the spatial distribution of young stars in the Perseus, Orion~A, 
Orion~B, and MonR2 molecular clouds. After subtracting a semi-empirical model 
of the field star contamination from the observed star counts, stellar surface 
density maps are used to identify compact clusters and any stellar population 
found more uniformly distributed over the molecular cloud. Each cloud contains
between 2 to 7 clusters, with at least half of the cluster population found 
in a single, rich cluster. In addition, a distributed stellar population is 
inferred in the Orion~A and MonR2 molecular clouds within the uncertainties 
of the field star subtraction with a surface density between 
0.013 - 0.083\sqamin. Sensitivity calculations suggest, however, that the 
number of stars in the distributed population may be underestimated by a 
factor of 2 or more if stars have been forming with a Miller-Scalo IMF at a 
constant star formation rate for longer than 10~Myr. After considering the 
possible evolutionary status of the distributed population, the global star 
formation efficiency implied by the sum of the distributed and cluster 
populations ranges between 1-9\% among the four clouds. The fraction of the 
total stellar population contained in clusters for the nominal extinction 
model ranges from \about 50-100\% if the distributed population is relatively 
young ($<$~10~Myr), to \about 25\%-70\% if it is relatively old (\about 
100~Myr). The relatively high fraction of stars contained in clusters 
regardless of the age of the distributed population, in conjunction with the 
young ages generally inferred for embedded clusters in nearby molecular 
clouds, indicates that a substantial fraction of the total stellar population 
in these regions has formed within the past few million years in dense 
clusters. This suggests that either the star formation rate in each these 
clouds has recently peaked if one assumes clouds have ages $>$ 10~Myr, or 
molecular clouds are younger than typically thought if one assumes that the 
star formation rate has been approximately constant in time.

\end{abstract}


\section{Introduction}

Nearby star forming regions display a continuum of properties, ranging from 
isolated young stellar objects and loose ``aggregates'' \citep{Gomez93,Strom93}
to dense clusters containing hundreds if not thousands of stars within parsec 
sized regions \citep{Lada91b,Carp97,Lynne98}. The diverse range of observed
stellar environments reflect the various physical processes that initiate star 
formation over the lifetime of molecular clouds. In principle, the predominant 
mechanisms that lead to star formation can be inferred by reconstructing the 
star formation rate in molecular clouds as a function of time and stellar 
mass. While a comprehensive picture of how star formation evolves in time
requires obtaining masses and ages for the individual stars that have formed 
within a cloud, constraints on the star formation history can nonetheless be 
obtained from a census of the embedded stellar population to establish the 
number of stars that have formed and where in the cloud they tend to be 
located.

The stellar population associated with a cloud is most often identified using 
large scale surveys for an emission feature commonly associated with 
pre-main-sequence stars, such as far-infrared emission (e.g. Beichman 
\etal~1986), near-infrared excesses (e.g. Cambr\'esy \etal~1998), H$\alpha$ 
(e.g. Nakano, Wiramihardja, \& Kogure~1995), or X-rays (e.g. Walter \etal~1988).
While such studies generally produce reliable 
catalogs of young stellar objects, they may not provide a complete census of 
the stellar population in molecular clouds. Surveys for far-infrared emission, 
near-infrared excesses, and H$\alpha$ in particular generally detect 
pre-main-sequence objects surrounded by an optically thick circumstellar disk 
or envelope. The evolution of circumstellar material as a function of stellar 
age and mass remains poorly quantified, which makes it difficult to infer the 
extent of the stellar population not currently in this phase of stellar 
evolution and not detectable by these diagnostics. Indeed, X-ray observations, 
which probe stellar chromospheric activity and can readily detect stars as old 
as \about 100~Myr in nearby (\aboutless 200~pc) star forming regions, have 
detected a widespread population of objects in and around molecular clouds. 
The interpretation of the X-ray results remain controversial, though, as the 
spatially extended X-ray population may represent older pre-main-sequence 
stars that have dispersed from molecular clouds, stars that formed in 
surrounding cloudlets \citep{Feigelson96}, or young main sequence stars that 
have formed in the solar neighborhood in the past 100~Myr 
\citep{Briceno97,Favata97}. Furthermore, X-ray observations in the pre-Chandra 
era lack the required combination of sensitivity, coverage, and angular 
resolution to provide an accurate census of the stellar population contained 
in both isolated star forming regions and dense clusters over entire molecular
clouds.

An alternative approach to infer the underlying stellar population in 
molecular clouds is to statistically determine the excess number of stars 
observed toward the cloud relative to the expected field star contribution. 
Such star count techniques are best suited for molecular clouds located at 
high Galactic latitudes where the field star contamination is less severe, and 
for near-infrared wavelength surveys where embedded stars are more readily 
detected. While near-infrared star count analyses have the obvious limitation 
that individual young stellar objects cannot be uniquely identified for 
further study, they represent a powerful probe of the stellar population in 
that near-infrared emission generally originates from the stellar photosphere 
(except for the heavily accreting objects). The intrinsic brightness of a star 
then depends foremost on the stellar mass (and age for pre-main-sequence 
objects), and the parameter space probed by a set of observations can be 
estimated with the aid of pre-main-sequence evolutionary models. Further,
current observations are generally sensitive to stars spanning a broad range 
of masses and ages, including objects in nearby molecular clouds with masses 
below the hydrogen burning limit, and have the angular resolution needed to 
resolve many of the stars in dense clusters. Thus near-infrared star count 
studies have the potential obtain a more unbiased census of the embedded 
stellar population in molecular clouds.

The first extensive near-infrared observations to statistically characterize 
the spatial distribution of young stars in a molecular cloud was the \KBAND\ 
band survey of Orion~B (L1630) by Lada \etal~(1991; see also Li, Evans, \& 
Lada~1997). Surveys of other regions with similar aims have since been
conducted for Orion~A (L1641; Strom, Strom, \& Merrill~1993), \NGC2264 
\citep{Lada93,Piche93}, the Rosette \citep{Phelps97}, \NGC1333 \citep{Lada96}, 
and Ophiuchus \citep{Strom95,Barsony97} among others (see recent reviews by 
Allen \& Hillenbrand~2000; Meyer \& Lada~2000; Clarke, Bonnell, \& 
Hillenbrand~2000). A synthesis of these results suggest that more than half of 
the stellar population in molecular clouds is contained in clusters with 
various sizes and densities \citep{Clarke00}. However, most near-infrared 
observations to date cover only a fraction of the total cloud area and nearly 
all have been predisposed toward regions within clouds known {\it a priori} to 
contain prominent star formation activity (and likely clusters). Thus current 
near-infrared observations do not accurately characterize the spatial 
distribution of stars over entire molecular clouds.

The 2 Micron All Sky Survey (\MASS) removes many of the limitations of prior 
observations and has provided the first sensitive, uniformly calibrated 
\JB, \HB, and \KB\ near-infrared survey over entire molecular clouds. In this 
paper, we analyze data from the \MASS\ Second Incremental Release to 
investigate the spatial distribution of young stars in the Perseus, Orion~A, 
Orion~B, and Monoceros R2 (MonR2) molecular clouds in order to determine the 
relative importance of cluster and isolated stellar systems. These clouds were 
chosen for this study since they are nearby with distances less than 1~kpc to 
maximum sensitivity to the low mass stellar population, are located at high 
Galactic latitudes ($|b| > 10$\arcdeg) to minimize field star contamination, 
and have available \thco\ maps that are required for the analysis presented 
here. The results from this study are presented as follows. In 
Section~\ref{data}, we review the characteristics of the \MASS\ data and 
select a subset of the point source catalog for analysis. In 
Section~\ref{analysis}, the near-infrared star counts toward the Perseus, 
Orion~A, Orion~B, and MonR2 molecular clouds are used to identify any cluster 
and distributed stellar populations. The implications of these results for the
star formation history of molecular clouds is discussed in 
Section~\ref{discussion}, and the conclusions are summarized in 
Section~\ref{summary}.

\section{\MASS\ Data}
\label{data}

The near-infrared data analyzed in this study were obtained from the \MASS\
Second Incremental Data Release. The observational goal of \MASS\ is to survey 
$>$95\% of the sky simultaneously in the \JB, \HB, and \KB\ bands. The data
are obtained by scanning a series of $8.5'\times6$\arcdeg\ long tiles aligned 
in declination. Each position in a tile is observed first with a short, 
51~msec integration to recover photometry for bright stars (between \about 
5-8\M), and then with a relatively long, 1.3~sec integration for deeper 
photometry. Adjacent images within a tile overlap such that each position on 
the sky is nominally observed 6 times. The pixel scale of the observations 
is 2\arcsec, but since overlapping images within a tile are designed to 
subsample the raw camera pixels, the coadded images produced in data reduction 
from the long integration frames have a pixel scale of 1\arcsec. The \MASS\ 
Second Incremental Release Point Source Catalog contains a list of positions 
and magnitudes for point sources in \about 47\% of the sky that have a 
photometric uncertainty $\le$ 0.155\M\ (i.e. a signal to noise ratio $\ge 7.0$)
in at least one of the three observed bands. This study analyzes a subset of 
the catalog that includes the Perseus, Orion~A, Orion~B, and MonR2 molecular 
clouds. At the time of the Second Incremental Release, \MASS\ data are 
available for the entire area of the MonR2 and Orion~A molecular clouds, 
\about 80\% of Perseus, and \about 70\% of Orion~B.  The galactic coordinates 
of these clouds and their adopted distances are summarized in 
Table~\ref{tbl:clouds}. 

The \MASS\ point source data need to be analyzed at magnitude limits bright
enough to ensure uniform completeness across the region studied. The magnitude 
thresholds for the three bands were set by computing as a function of 
magnitude the average signal to noise ratio and the fraction of stars that 
have a signal to noise ratio $\ge$ 10 for point sources between Galactic 
longitudes of 130\arcdeg\ and 250\arcdeg\ and Galactic latitudes of 
-50\arcdeg\ and +50\arcdeg. Based on this analysis, the faint magnitudes 
thresholds for analyzing the star counts were set at 15.8\M, 14.9\M, and 
14.3\M\ at \JB, \HB, and \KB\ band respectively. At these thresholds, the 
average signal to noise for the point sources is between 14-15, and greater 
than 99\%, 99\%, and 97\% of the sources at these magnitude limits have a 
photometric uncertainty corresponding to a signal to noise ratio $\ge$ 10. 
Stars brighter than \about 5\M\ are saturated even in the short integration
data, and as a conservative limit, it was required that stars be fainter than 
6.0\M\ in the appropriate band to be included in the analysis. At the galactic 
coordinates of the clouds studied here, the average source densities at the 
faint magnitude thresholds are \aboutless 2\sqamin, which implies a typical 
source separation of \aboutmore 40\arcsec. Thus source identification will 
only be confusion limited in the densest clusters which can have substantially 
higher surface densities than these average values.

Individual survey tiles may have poorer sensitivity than the adopted magnitude 
thresholds if spatial structure in the sky background from airglow emission 
added substantially to the sky noise (especially at \HB\ band) at the time of 
the observation or if the sky background was much higher than average 
(especially at \KB\ band). Data adversely affected by airglow were identified 
and removed in the present analysis by finding all tiles in the Second 
Incremental Release Scan Database in which the observed sky noise was 20\% 
higher than the expected noise computed from the sky background level. The 
percentage of the total number of tiles (\about 9400 total) removed due to 
this criteria amounted to 0\%, 8\%, 0.1\% at \JB, \HB, and \KB\ band 
respectively. To remove lower sensitivity data because of a high sky 
background level, tiles in which the average signal to noise ratio is less 
than 10.0 at the adopted magnitude thresholds were also discarded. The 
percentage of tiles removed by this criteria were 0.14\%, 0.07\%, and 0.25\% 
at \JB, \HB, and \KB\ band.

Finally, some point sources identified around bright stars ($m$ \aboutless 
9.5\M) were omitted when drafting the Second Incremental Release Catalogs so 
that image artifacts from bright objects would not reduce the data reliability.
The masked areas are primarily due to diffraction spikes and circular 
regions centered on bright saturated stars. The dimensions of these masks
depend on the magnitude of the bright star and effectively create holes in the 
star count maps. Using the quantitative description of these masks provided in 
the \MASS\ Explanatory Supplement, the fraction of each pixel in the star 
count map that has been partially or completely masked was computed as a 
function of magnitude and was used to correct the star counts maps on a pixel 
by pixel basis.

\section{Analysis}
\label{analysis}

A \KB\ band stellar surface density map for the area between Galactic 
longitudes of 130\arcdeg\ and 250\arcdeg\ and Galactic latitudes of 
$-40$\arcdeg\ and $+40$\arcdeg\ is shown in Figure~\ref{ksqrt}. The surface
density map was generated by binning stars with \KB\ band magnitudes between 
6.0\M\ and 14.3\M\ in $5'\times5'$ cells. The Galactic Plane is clearly 
visible as a band of stars extending across the center of the image. The 
location of the Perseus, Orion A, Orion B, and MonR2 molecular clouds are 
labeled in Figure~\ref{ksqrt}, but because of the low \KB\ band extinction, 
the outline of the clouds are only faintly seen in absorption against the 
backdrop of field stars. To gain a better perspective on the location of 
various molecular clouds, Figure~\ref{jk} presents a map of the average \JK\ 
stellar color for the same region shown in Figure~\ref{ksqrt}. Large \JK\ 
colors, represented by darker gray scales in Figure~\ref{jk}, are attributed 
to background field stars and embedded young stellar objects reddened by dust 
associated with molecular clouds. The Taurus, Perseus, Orion~A, Orion~B, and 
MonR2 molecular clouds are clearly visible in Figure~\ref{jk}, as well as 
numerous clouds in the Galactic Plane. The outline of the clouds agree well 
with that inferred from optical star counts \citep{Cam99} and far-infrared 
emission \citep{Reach98}. 

The observed star counts toward a molecular cloud consist of unreddened 
foreground field stars, background field stars reddened by dust within the 
cloud, and young stars associated with the cloud itself. The spatial 
distribution of young stars in the cloud can then be obtained by removing the 
field stars from the star count map. Ideally one would like to specifically
identify the individual young stars so that they can be furthered studied. 
However, near-infrared photometry alone cannot uniquely distinguish field 
stars from the stellar population associated with the cloud. Stars with 
relatively blue near-infrared colors can either be foreground field stars, 
background field stars appearing through holes in the cloud, or young stars on
the surface of the cloud. Similarly, red stars can either be background 
reddened field stars or pre-main-sequence objects embedded in the cloud. The 
degeneracy between the field and embedded stellar population can be only
partially reduced by considering both the magnitudes and colors. Therefore, 
instead of attempting to uniquely identify which stars are 
associated with the cloud, the embedded stellar population was inferred 
statistically by first constructing a semi-empirical model of the expected 
field star surface density toward the cloud, and then subtracting the field 
star model from the observed stellar surface density map. In principle, the 
star count analysis can be done by simultaneously analyzing the five 
observable quantities for each star (i.e. Galactic longitude and latitude and 
the observed \JB, \HB, and \KB\ band magnitudes). In the analysis conducted 
here, the three near-infrared bands are analyzed separately to determine if 
each band provides a consistent picture of the embedded stellar population. 
The following sections describe how the semi-empirical field star model was 
generated using \MASS\ observations of regions away from the molecular clouds 
and published \thcoj\ maps as a tracer of the cloud extinction. The field star 
subtracted stellar surface density map is then used to identify stellar 
clusters and any stars distributed more uniformly throughout the molecular 
cloud.

\subsection{Semi-Empirical Field Star Model}
\label{model}

The expected number of field stars toward the molecular clouds in the absence
of extinction from the cloud itself was estimated using \MASS\ observations
of regions outside the molecular cloud boundaries. This was accomplished by 
fitting Legendre polynomials as a function of Galactic longitude and latitude 
to \JB, \HB, and \KB\ surface density maps for the region between $\ell = 
130$\arcdeg\ to 240\arcdeg\ and $b = -35$\arcdeg\ to $-7$\arcdeg\ binned in
$5'\times5'$ cells (see Figure~\ref{ksqrt}). This bin size was chosen so that 
large scale gradients in the field star counts are well resolved while 
maintaining a manageable image size. A global fit was performed to the large 
scale surface density maps as opposed to fitting a localized region around 
each cloud since several blocks of \MASS\ tiles were not available at the time 
of this study, especially near the Orion~B molecular cloud. Without these 
tiles, a localized polynomial fit is poorly constrained near the cloud 
boundaries. Lines of sight in the surface density map that pass through an 
obvious stellar cluster (see Figure~\ref{ksqrt}) or that intercept a molecular 
cloud as traced by molecular emission 
\citep{Mad86,Dame87,Bally87,Miesch94,Padoan99} or red \JK\ colors (see 
Figure~\ref{jk}) were excluded from the fit. At least a 0.5\arcdeg\ border
around each cloud was also excluded which prevented any potential young stars 
surrounding the clouds from influencing the polynomial fit. The remaining 
\about 180,000 pixels in each surface density map were then fitted with 
Legendre polynomials.
The order of the polynomial fit was increased until the systematics in the 
residuals as a function of position were no longer evident. A seventh
order polynomial fit in both longitude and latitude were ultimately used.
The results from the polynomial fit are examined in Figure~\ref{fit}, which 
shows as a function of Galactic latitude the observed mean stellar surface 
density, the mean residuals after subtracting the polynomial fit, and the RMS 
in the residuals for each band. The RMS of the residuals varies between \about 
0.1-0.3\sqamin\ depending on the band and the Galactic latitude, and is 
consistent with that expected from Poisson statistics within the $5'\times5'$ 
bin sizes.

The polynomial fit described above was interpolated to yield the expected 
surface density of field stars as a function of Galactic longitude and 
latitude toward the molecular clouds in the absence of extinction from the
cloud itself. The uncertainties associated with interpolating the polynomial 
fit were assessed by performing a similar fit to star counts at the same
Galactic longitude but at positive latitudes where are there essentially no
clouds (see Figure~\ref{jk}). The molecular clouds masks used in fitting the 
star counts at negative latitude were also used at positive latitudes in
in order to preserve the geometry and number of masked regions. The accuracy 
of the interpolated polynomial fit was assessed by subtracting the fit from 
the observed star counts in the masked regions and computing the RMS of the 
residuals in areas 6~deg$^2$ in size, which is a typical area of the larger 
clouds analyzed here. It was found that the average difference over this sized 
region between the interpolated fit and the observed star counts is 
\aboutless 0.003\sqamin\ in each band with a RMS deviation of \about 
0.002\sqamin.

Dust within the molecular clouds obscures background field stars and depresses 
the total field star counts relative to the interpolated polynomial fit. The 
number of field stars that are obscured depends on the extinction along the 
line of sight and the frequency distribution of magnitudes for the background 
field star population, as foreground field stars will not be further reddened 
by the cloud. The extinction as a function of position through the molecular 
clouds was estimated using published \thco\ maps of the MonR2, Orion~A, 
Orion~B, and Perseus molecular clouds kindly provided by J. Bally (see Padoan 
\etal~1999; Miesch \& Bally~1994; Bally \etal~1987). These maps have a 
full-width-at-half-maximum beam size of 100\arcsec\ with a sampling of 
60\arcsec. While the molecular gas will contain substructure within the 100$''$
beam, these maps resolve the large scale structural features present in these
clouds and should accurately trace the global spatial distribution of 
extinction.

Assuming that the \thcoj\ emission is optically thin and in Local 
Thermodynamic Equilibrium, the \thco\ column density can be estimated from the 
J=1-0 integrated intensity using the formula
\begin{equation}
\label{extinction}
    \rm{N(^{13}CO)} = 4.57\times10^{13}\:T_{ex}\:e^{5.287/T_{ex}} \int 
        {T_{mb}\:\Delta v \over K\:km s^{-1}}\;\rm{cm}^{-2},
\end{equation}
where $T_{ex}$ is the excitation temperature and $\int\rm{T_{mb}\Delta v}$ the 
integrated \thcoj\ intensity. The derived \thco\ column densities will be in 
error by less than a factor of 2 by assuming $T_{ex}$=10~K as long as the 
actual excitation temperature is less than 30~K. Such high excitation 
temperatures are expected only toward the embedded OB stars which occupy a 
small fraction of the total cloud area in any of these regions. The \thco\
columns densities can be converted into visual extinctions using the empirical 
$N(^{13}CO)$--$A_V$ correlations that has been observed in many clouds 
\citep{Frerking82,BC86,Langer89,Lee94,Lada94,Hayakawa99}. These studies
have shown that the ratio of $N(^{13}CO)$ to $A_V$ is typically within the 
range of \about $1.5-2.5\times10^{15}$~\cmmag\ and that \thco\ is detectable 
only when the visual extinction reaches a threshold value of \about 0.5-1.5 
magnitude due to photodissociation of the molecules at low column densities 
\citep{Frerking82}. To determine the sensitivity of the field star model on 
the derived extinctions, three different set of assumptions were adopted to 
convert the \thco\ column densities into visual extinctions. The ``nominal'' 
extinction model assumes a $N(^{13}CO)/A_V$ ratio of 
$2.18\times10^{15}$~\cmmag\ as derived by \citet{Lada94} and that the 
extinction needed to detect \thcoj\ is $A_V=1$\M. This calibration was adopted
since it has the best statistical accuracy among published observations and
the parameters are in the middle of values generally observed in molecular 
clouds. Similarly, a ``low'' 
extinction model was derived by assuming a $N(^{13}CO)/A_V$ ratio of 
$2.5\times10^{15}$~\cmmag\ and a visual extinction threshold of 0.5\M, and a 
``high'' extinction model was obtained by assuming a $N(^{13}CO)$/$A_V$ ratio 
of $1.5\times10^{15}$~\cmmag\ and a visual extinction threshold of 1.5\M. The 
low and high extinction models adopt parameters at the extrema of the observed
values and are assumed to represent the plausible range of visual extinction
and are not formal 1$\sigma$ confidence intervals. The masses of the clouds 
under these three extinction assumptions are summarized in 
Table~\ref{tbl:clouds}.

The observed correlation between \thco\ column density and visual extinction 
is valid for $A_V \le5$\M\ \citep{Lada94}. For $A_V$ \aboutmore 10\M, the 
\thcoj\ emission is saturated and is not a reliable tracer of the extinction. 
The \thco\ maps used here have a 3$\sigma$ detection level corresponding to a 
visual extinction of \about 0.2\M\ (not including the visual extinction 
threshold) for the nominal extinction model. The fraction of the Perseus, 
Orion~A, Orion~B, and MonR2 cloud area with an inferred visual extinction 
$\le$ 5\M\ is 95\%, 88\%, 92\%, and 98\% respectively. Therefore, \thcoj\ 
emission should accurately trace the extinction over nearly the entire 
cloud area. 

The last step in constructing the field star model is to estimate the fraction 
of background field stars that become obscured by dust within the molecular 
clouds. This fraction was estimated using the extinction maps described above 
and the \citet{Wainscoat92} Galactic star count model. This model agrees with 
the observed star counts to within 15\% for high Galactic latitude regions 
($|b| > 5$\arcdeg) that do not contain additional reddening along the line of 
sight from molecular clouds. Since Perseus, Orion~A, Orion~B, and MonR2 are 
all located at Galactic latitudes between $-20$\arcdeg\ and $-10$\arcdeg, we 
expect that this model should provide an accurate estimate of the fraction of 
field stars that are foreground to the molecular clouds. The 
\citet{Wainscoat92} model predicts that the fraction of the total field star 
counts contributed by foreground field stars ranges from 0.05-0.08 in Perseus 
to 0.2-0.4 in MonR2 in the \JB, \HB, and \KB\ bands. The predicted absolute 
level of the foreground field star contamination is between 0.04\sqamin\ 
(Perseus) and \aboutless 0.3\sqamin\ (MonR2).

For each line of sight toward the molecular clouds with a \thcoj\ detection,
the expected number of field stars were computed using the Wainscoat star 
model with and without the \thco\ derived extinction added at the appropriate 
cloud distance. The fraction of the field star population in this model that
are reddened to magnitudes outside the adopted thresholds (see 
Section~\ref{data}) was then computed, and the expected field star 
contribution inferred from the polynomial fits was reduced by the same 
fraction for that line of sight. Thus the total number of unreddened field 
stars is still determined by \MASS\ observations, and only the relative 
fraction of field stars obscured by extinction from dust in molecular clouds is 
model dependent. In performing the calculations, we adopted the interstellar 
reddening vector from \citet{Cohen81} and the cloud distances summarized in
Table~\ref{tbl:clouds}.

Uncertainties in the molecular cloud distances and the three dimensional
structure of the clouds contribute to uncertainties in the field star model 
since extinction from the molecular cloud is applied only to the background 
field stars. The distance dependent uncertainties on the field star 
subtraction were quantified using the \citet{Wainscoat92} star count model. 
Adopting an average visual extinction of 3\M\ as implied by the \thco\ maps, 
the $K$ band field star surface density computed from the Wainscoat model will 
vary by \about $\pm$ 0.002\sqamin\ for Perseus, $\pm$ 0.004\sqamin\ for Orion~A
and Orion~B, and $\pm$ 0.014\sqamin\ for MonR2 by changing the cloud distance 
by $\pm$ 20\% from the nominal assumed value. The corresponding uncertainties 
at \JB\ and \HB\ bands are higher by a factor of \about 2-3. The uncertainties 
are largest for MonR2 since it is the most distant cloud analyzed here and the 
foreground field stars contain a greater fraction of the total field star 
counts than toward Perseus, Orion~A, or Orion~B. In addition, since molecular 
clouds are three dimensional structures, there may be systematic variations in 
the distances to different parts of the cloud. For the Orion molecular clouds
in particular, \cite{Brown94} argue that the near edge of the Orion~A and 
Orion~B molecular clouds is at a distance of \about 320\pc, and the far edge 
is at a distance of \about 500\pc. The difference in the field star model 
between these two distances for a visual extinction of 3\M\ is \about 
0.006\sqamin.

\subsection{Field Star Subtracted Surface Density Maps}
\label{density}

The stellar population associated with the cloud can be inferred by subtracting
the field star model from the observed star counts. Instead of using the 
binned star count maps that were produced for the field star polynomial fit,
(e.g. Figure~\ref{ksqrt}), stellar surface density maps were re-generated for 
the molecular clouds from the observed star counts using adaptive kernel 
density estimation \citep{Silverman86}. In 
this method, each star is represented by a kernel function, such as a gaussian, 
where the width of the kernel depends on the local stellar density as 
described below. The sum of the individual kernel functions yields the stellar 
density map for the cloud. The advantage of adaptive kernel density estimation 
compared to a fixed kernel size or a binned star count map is that the 
resolution of the map varies with the local stellar density which enables 
compact clusters to be more readily identified.

The kernel width for each star was computed using the procedure outlined 
in \citet{Silverman86}. First, a pilot stellar density map, $\tilde 
f(\ell,b)$, was constructed using the standard kernel density estimate as
\begin{equation}
   \tilde f(\ell,b) = \sum_{i=1}^n K_i(h,\ell,b),
\end{equation}
where $h$ is the bandwidth that controls the amount of smoothing, $K$ is the 
kernel function, and $n$ is the number of stars. The results are not sensitive 
to the choice of the Kernel function, and a radially symmetric gaussian kernel 
was adopted such that
\begin{equation}
   K_i(h,\ell,b) = {{\rm cos}\:b_i\over2\pi\:h^2}\;
      e^{-{(\ell-\ell_i)^2\:{\rm cos}^2b_i\over2h^2}}\;
      e^{-{(b-b_i)^2\over2h^2}}.
\end{equation}
For the pilot density estimate, the bandwidth, $h$, was fixed at the angular 
size corresponding to 0.25~pc in anticipation that clusters and their 
associated dense molecular cores have typical sizes of \about 1\pc. A local 
bandwidth, $\lambda_i$, was then computed for each star as
\begin{equation}
\lambda_i = [\tilde f(\ell_i,b_i)/g]^{-\alpha},
\end{equation}
where $g$ is the geometric mean of $\tilde f(\ell_i,b_i), i=1,n$. Following 
\citet{Silverman86}, $\alpha$ was set to be 0.5. Thus in regions of low 
stellar surface density such that $\tilde f(\ell,b)$ is small, the local 
bandwidth, $\lambda$, is relatively large and the data are heavily smoothed. 
Conversely, in regions of high density such that $\tilde f(\ell_i,b_i)$ is 
large, $\lambda$ is relatively small and the data are smoothed less. The 
adaptive kernel density estimate was then computed as
\begin{equation}
\hat f(\ell,b) = \sum_{i=1}^n K_i(h\lambda_i,\ell,b).
\end{equation}
As for the pilot density estimate, the bandwidth, $h$, was chosen to be 
0.25~pc. The expected number of field stars for each pixel in the map (see 
Section~\ref{model}) was then subtracted from the adaptive kernel surface 
density image to obtain the stellar population associated with the cloud.

\subsection{Stellar Clusters}
\label{clusters}

\KB\ band surface density maps for the Perseus, Orion~A, Orion~B, and MonR2 
molecular clouds are presented in 
Figures~\ref{perseus_kernel}--\ref{monr2_kernel} along with corresponding
\IRAS\ 60\micron\ images, \thcoj\ integrated intensity maps 
\citep{Bally87,Miesch94,Padoan99}, and an image of the average \JK\ color for
points sources in the \MASS\ Second Incremental Release. Based on the extent
of the molecular clouds indicated by the \thco\ images, the \MASS\ 
observations at the time of this study were not available for the southern
portion of the Perseus molecular cloud and the northern third of the Orion~B
molecular cloud. Further, the Orion~A \thco\ map from \citet{Bally87} does not 
encompass the full extent of the molecular emission near $\ell=214^\circ$ 
(see Nagahama \etal~1998).

Stellar clusters were identified by forming closed contours at the $2\sigma$ 
level above the local background level in the field star subtracted density 
maps. The noise in the star counts, $\sigma$, was determined by assuming 
Poisson statistics for the field stars at the appropriate latitude and 
longitude. Clusters are defined as closed contours that have a peak stellar 
surface density $\ge 6\sigma$ and a total number of stars within the closed 
contour $\ge 4\sigma$ with respect to the expected field star population. 
These thresholds are arbitrary, and were selected using lines of sight away 
from the molecular cloud as a control field that should be dominated by field 
star population and contain few if any clusters. While the 6$\sigma$ peak 
density requirement biases the algorithm from finding extended, low density 
clusters, the resulting list of the clusters should be fairly reliable.

The clusters identified in the field star subtracted density map are listed in 
Table~\ref{tbl:clusters} along with the number of cluster members, the angular 
area within the 2$\sigma$ contour, and any associations with known star 
forming regions. All of the identified clusters are located at least partially
in projection against the molecular cloud boundaries. Three of the clusters in 
Orion~A are outside the spatial extent of the \thco\ image. As discussed 
further below, lower limits to the cluster membership are provided for a few 
clusters since part of the cluster area has been masked in the \MASS\ Point 
Source Catalog due to 
neighboring bright stars. The clusters are labeled in the surface density maps 
in Figures~\ref{perseus_kernel}--\ref{monr2_kernel}, and contour maps of the 
\KB\ stellar surface for each cluster are presented in Figure~\ref{contours}. 
The major axis of the cluster measured at the 2$\sigma$ contour level ranges 
from \about 1~pc to \about 16~pc for the Orion Nebula Cluster (ONC).
$J-H$ vs. $H-K_s$ color-color diagrams for stars within the projected area of 
each cluster are shown in Figure~\ref{jhhk}. Each cluster contains a number of 
stars with red near-infrared colors, supporting the notion that these clusters 
are indeed embedded within the molecular clouds. 

Most of the clusters shown in Figure~\ref{contours} have been previously
identified and studied. In the Perseus molecular cloud, the \IC348 cluster 
has been the subject of several photometric and spectroscopic investigations 
\citep{Lada95,Herbig98,Luhman98,Luhman99}. In the Orion~A molecular cloud, the
ONC \citep{JW88,Prosser94,Ali95,Lynne97} is by the far largest, richest cluster 
identified here and extends for \about 2\arcdeg\ (16\pc) along a filament of 
molecular gas prevalent in molecular line \citep{Bally87,Tatematsu98,Nag98}
and dust continuum \citep{Lis98,Johnstone99} images. The boundaries of the ONC 
as defined here include the L1641N cluster \citep{Strom93,Chen93b,Chen94} and 
a couple of stellar density enhancements associated with \IRAS\ point sources 
\citep{Chen94}. In the Orion~B molecular cloud, \NGC2024\ 
\citep{Lada91b,Comeron96,Meyer96} and \NGC2068 \citep{Lada91b} are identified 
as clusters, although \NGC2068 is at the edge of the density map and has not 
yet been completely imaged by \MASS. Finally, the major cluster identified in 
the MonR2 molecular cloud is the MonR2 cluster itself 
\citep{Aspin90,Howard94,Carp97}. 

The number of cluster members listed in Table~\ref{tbl:clusters} are lower 
limits to the total cluster membership in these clouds. As with any survey, 
the completeness of the cluster membership is limited by the ability to 
resolve stars in crowded regions, the sensitivity of the observations to stars 
of various ages and masses, the high extinction in many clusters that may 
obscure some stars, and the difficulties in detecting point sources against 
bright nebular backgrounds that often accompany young star forming regions. 
Further, as a feature of the \MASS\ Second Incremental Release Point Source 
Catalog, part of the cluster area in \IC348, \LDN1641~C, the ONC, and 
\NGC2024 have been masked out due to the presence of nearby bright stars (see 
Figure~\ref{contours}). 
The fraction of the cluster area masked out in the \IC348, \LDN1641~C, and ONC 
clusters is estimated to be \about 10\%, 5\%, and 5\% respectively. The masked 
region in the ONC includes the dense Trapezium region that contains \about 600 
hundred stars brighter than $K$=14\M\ within the central $5'\times5'$ of the 
cluster \citep{MS94,Lynne00}. To assess the fraction of the cluster population 
masked out in \NGC2024, the \MASS\ observations were compared with the 
\NGC2024 $K$ band survey by \citet{Lada91b}, which has comparable resolution 
and sensitivity as the \MASS\ observations. They identified 309 stars within 
the cluster boundaries, compared to $>$201 stars found here.

The density and molecular cloud maps presented in 
Figures~\ref{perseus_kernel}-\ref{monr2_kernel} show that the \MASS\ 
observations at the time of this study were not complete toward the Perseus 
and Orion~B molecular clouds. The prominent star forming regions 
contained within these unscanned areas include the \NGC1333 cluster in 
Perseus \citep{Aspin94,Lada96,Aspin97}, the \NGC2071 cluster in Orion~B
\citep{Lada91b}, and the remaining part of \NGC2068 in Orion~B.
Previous observations with comparable sensitivity and resolution as the \MASS\
data suggest that the \NGC1333 cluster contains \about 94 stars \citep{Lada96} 
and \NGC2071 \about 105 stars \citep{Lada91b}. The \NGC2068 cluster contains 
\about 147 stars in addition to those already found identified \citep{Lada91b}. 
Since the smallest cluster identified with the \MASS\ data contains 15 stars, 
each of these regions would almost certainly have been identified as a cluster 
if they have been included in the \MASS\ Second Incremental Release. 
Several star forming regions in the Orion clouds found in previous studies did 
not meet the cluster definition adopted here, including \NGC2023, HH34, 
V380~Ori, the Cohen-Kuhi group, KMS 35, KMS 36, and additional low luminosity 
\IRAS\ sources \citep{Lada91b,Strom93,Chen94}. These star forming regions 
contains between 5 and 34 stars each, which is near the limit of the smallest
clusters define in this study. The total cluster membership in Orion~A and 
Orion~B would not change appreciably if these regions were formally defined as 
clusters in this study.

Given the high extinction that is generally associated with star 
forming regions, deeply embedded, rich clusters (\aboutmore 100 stars) could 
exist that remain undetected at near-infrared wavelengths. Any such clusters 
should appear in the \IRAS\ 60\micron\ image which can probe deeper into 
molecular clouds than \KB\ observations. In Perseus, the brightest \IRAS\ 
sources besides \IC348 and \NGC1333 are associated with \LDN1448, \LDN1455, 
and a ring of bright far-infrared emission southwest of \IC348 that is 
possibly related to a \hii\ region \citep{Andersson00}. The luminosities 
of these \IRAS\ point sources are \aboutless 20\lsun\ \citep{Ladd93}, which is 
too low to contain a significant cluster of young stars unless the stellar 
masses are strongly biased toward low mass objects. In Orion~A and Orion~B, 
each of the bright far-infrared emitting regions corresponds to a stellar 
cluster and therefore it is unlikely that a rich cluster remains unidentified 
in these clouds. In MonR2, three bright \IRAS\ sources, corresponding to 
reflection nebula VDB~70, VDB~72 (\NGC2182), and VDB~74 \citep{VDB66}, have
no apparent clusters in the \MASS\ data. The relatively blue stellar colors 
and the weak \thco\ emission indicate that the molecular gas associated with 
these nebula is diffuse and cannot contain a deeply embedded cluster. Two 
additional bright \IRAS\ sources (06124-0621 and 06128-0624) are present in 
the northeast corner of the MonR2 molecular cloud that are coincident with 
moderately bright \thco\ emission. Part of the \MASS\ point source data in 
this region has been masked due to a neighboring bright star. Visual 
inspection of the \MASS\ Image Atlas suggests that \IRAS\ 06128-0624 may be 
associated with a cluster of \about 20 stars, but the low far-infrared 
luminosity of both sources (\aboutless 30\lsun) suggests that neither \IRAS\ 
source is associated with a rich cluster. Therefore, it it seems likely that 
the richest clusters in each of these molecular clouds have been identified 
from \MASS\ and published observations.

In summary, the total number of cluster members identified in the \MASS\ data 
after field star subtraction is $>$299, $>$1992, 246, and 543 for Perseus, 
Orion~A, Orion~B, and MonR2, respectively. If we include observations from 
the literature to take into account clusters not included in the \MASS\ Second 
Incremental Release (\NGC1333 and \NGC2071), clusters that have been only 
partially mapped with \MASS\ so far (\NGC2068), and clusters that have been 
partially masked because of neighboring bright stars (\NGC2024), the total 
number of stars in clusters increases to 393 for Perseus and 606 for Orion~B. 
Corrections to the total stellar population in \IC348, \LDN1641~C, and the ONC 
due to masked out regions are not accurately known and have not been applied. 
The number of clusters ranges from 2 in Perseus to 7 in Orion~A, and at least 
half of the cluster stellar population in each cloud can be attributed to a 
single cluster. The fraction of the molecular cloud surface area as defined 
by \thcoj\ emission occupied by the clusters (including the cluster properties 
incorporated from the literature) is \about 3\%, 12\%, 7\%, 3\% in Perseus, 
Orion~A, Orion~B, and MonR2 respectively.

\subsection{Distributed Stellar Populations}
\label{distributed}

Stars associated with the molecular cloud but located outside the 
boundaries of the stellar clusters are defined as the distributed stellar 
population. The distributed population may include stars that formed in 
isolation, stars that formed in clusters in the past but have since dispersed 
into the 
molecular cloud, small groups of stars not meeting the adopted surface density 
requirements to be identified as clusters, and stars that are associated with 
identified clusters but lie beyond the adopted cluster boundary. The angular 
extent of the distributed population is unknown, so it is assumed that the 
stars may be located anywhere where \thcoj\ emission is detected as shown in 
Figures~\ref{perseus_kernel}-\ref{monr2_kernel}. 

The signature of a distributed population is a positive average stellar 
surface density toward lines of sight outside the cluster boundaries but 
within the molecular cloud after subtracting the field star contribution to
the observed star counts. Histograms 
of the field star subtracted, \JB, \HB, and \KB\ stellar surface densities 
toward the Perseus, Orion~A, Orion~B, and MonR2 molecular clouds are shown in 
Figures~\ref{perseus_hist}-\ref{monr2_hist} for the nominal extinction model.
The open histograms are for all lines of sight toward the molecular clouds, 
and the hatched histograms are for lines of sight within the cloud that 
do not intercept a cluster. Each of the histograms are similar in that they
peak at a surface density of \about 0\sqamin\ with a tail toward higher values.
The high surface density tails in the histograms are generally associated with 
clusters, although the \JB\ and \HB\ histograms for Orion~A do contain a few 
lines of sight with high surface densities ($> 1$\sqamin) that have been 
assigned to the distributed population. These lines of sight are regions 
around bright stars in the ONC that were masked out at \KB\ band where the 
cluster area was defined, but not at \JB\ and \HB\ bands where a smaller 
region was masked out. For each cloud, the histograms extend to negative 
surface densities. Negative surface densities for individual pixels are 
expected since Poisson fluctuations in the star counts implies that the number 
of field stars will be less than the expected average field star surface 
density for some lines of sight, and exceed the field star model for others. 
The number of lines of sight with a negative surface density and the width of 
the histograms in Figures~\ref{perseus_hist}-\ref{monr2_hist} depend on the 
kernel size used to generate the surface density maps and the surface density 
of the field star population. Since the nominal kernel width is set to a 
constant linear size of 0.25\pc, the angular size of the kernel will be 
smallest for the more distant cloud, MonR2, and largest for the nearest cloud, 
Perseus. The Perseus star count map will then be more heavily smoothed and 
have smaller fluctuations in the observed star counts about the model field 
star population than MonR2, as observed.

The mean surface density of the distributed population was obtained by 
averaging the field star subtracted surface density map for regions outside
the cluster boundaries but within the \thco\ map area as represented 
graphically by the hatched histograms in 
Figures~\ref{perseus_hist}-\ref{monr2_hist}. These mean surface densities are 
summarized in Table~\ref{tbl:distributed} for each band and the three 
extinction models. The formal 1$\sigma$ uncertainties in the mean stellar 
surface densities computed from the distribution of points in the hatched 
histogram are \about 0.005\sqamin\ in each band for Perseus, Orion~A, 
and Orion~B, and \about 0.009\sqamin\ for MonR2. By comparison, the range
in the surface densities due to the uncertainties in the extinction model
are \about $\pm$ 0.10, 0.05, and 0.03\sqamin\ at \JB, \HB, and \KB\ 
respectively (see Table~\ref{tbl:distributed}). As mentioned in 
Section~\ref{model}, however, the uncertainties due to the extinction model 
are not formal 1$\sigma$ confidence intervals but represent the plausible range
of values based upon converting the \thco\ integrated intensity into visual
extinction. Thus compared to the other field star model uncertainties 
discussed in Section~\ref{model}, the largest source of uncertainty in the 
mean stellar surface density for the distributed population is converting the 
\thco\ integrated intensity to visual extinction. 

The range of possible surface densities for the distributed population from 
the three extinction models is shown graphically in 
Figures~\ref{perseus_cuts}-\ref{monr2_cuts}. These figures show the field star 
subtracted surface density averaged over Galactic latitude for each cloud as a 
function of Galactic longitude for the nominal extinction model. The 
surface density of the distributed population for the nominal extinction model 
is shown by the solid horizontal line, and the dashed lines show the 
corresponding surface densities derived from the low and high extinction 
models. The top panel in each figure shows the average visual extinction also
averaged over Galactic latitude. The narrow peaks in the surface density at 
discrete Galactic latitudes represent the compact stellar clusters, although 
the large angular extent of the Orion Nebula Cluster is clearly visible. 
Figures~\ref{perseus_cuts}-\ref{monr2_cuts} indicate that within a given cloud,
the \JB, \HB, and \KB\ data give a consistent value for the distributed 
population surface density within the uncertainties of the field star 
subtraction. The most stringent constraints on the distributed population is 
provided at \KB\ band, which is the least sensitive of the three bands to the 
assumed extinction model. Therefore, the characteristics of the distributed 
population discussed below will be based primarily on the \KB\ band data.
Table~\ref{tbl:distributed} indicates that for the nominal extinction model,
the Perseus, Orion~A, and MonR2 molecular clouds contain a positive mean 
stellar surface density that is indicative of a distributed population, while
Orion~B has a negative surface density. For the low extinction model, the
Perseus molecular cloud has also a negative implied surface density for the
distribution population. A negative mean surface density for the distributed 
population is unphysical and indicates that too many field stars were 
subtracted from the observed star counts for that particular extinction model.
The number of stars contained 
in the distributed population over the entire cloud as mapped in \thco\ was 
computed by multiplying the average surface density by the cloud area (see 
Table~\ref{tbl:clouds}), including those regions not encompassed by the \MASS\
Second Incremental Release. The magnitude of the distributed population 
inferred from the \KB\ band observations ranges from zero stars in Orion~B to 
\about 730 in Orion~A. However, given the uncertainty in the extinction, a 
distributed population consisting of \about 790 stars cannot be ruled out even 
in Orion~B. These estimates for the number of stars in the distributed 
population refer only to stars that fall within the adopted magnitude 
thresholds. The fraction of the distributed stellar population that may exist 
but have magnitudes outside these thresholds is considered in 
Section~\ref{dis:sensitivity}.

The surface density of the distributed population can be compared with
previous observations of the Orion~A and Orion~B molecular clouds. 
\citet{Lada91b} found that the surface density for the distributed population 
in Orion~B of \about 0.007\sqamin\ for stars with $K < 13$\M\ over a 
0.7~deg$^2$ region. \citet{Li97} observed a 0.37~deg$^2$ region away from
known clusters at $J$, $H$, and $K$ band in Orion~B, and inferred that the 
surface density of stars in the distributed population with a near-infrared
excess in the $J-H$ vs. $H-K$ color-color diagram is $<$ 0.014\sqamin. Both of 
these results are consistent with values obtained here for Orion~B within the 
uncertainties of the field star subtraction. \citet{Strom93} analyzed
the stellar population in Orion~A based on a near-infrared imaging survey
encompassing a total of 0.77~deg$^2$ of the cloud area. In addition to 
identifying several clusters, they measured a distributed population of 
\about 1500 stars down to a 5$\sigma$ detection limit of $J=16.8$\M, implying 
a surface density for the distributed population of \about 0.54~stars\sqamin. 
By comparison, the maximum surface density for the distributed population in
Orion~A implied by the \MASS\ data is 0.099~stars\sqamin. The discrepancy 
between the \citet{Strom93} results and those obtained here are unclear. 
While \citet{Strom93} analyzed their data at a $J$ band magnitude limit 0.7\M\ 
fainter than adopted here, Figures~9 and 11 in that paper suggests that most 
of the distributed population they found should have $J$ band magnitudes 
brighter than 16\M\ and be detectable with the magnitudes limits adopted here. 
Given that the \MASS\ \JB, \HB, and \KB\ star counts analyzed here are based 
on more extensive observations and provide consistent results, we will use 
these data to characterize the distributed population in Orion~A.

\section{Implications for Global Star Formation Properties of Molecular Clouds}
\label{discussion}

A complete picture of the star formation history of molecular clouds requires
extensive spectroscopic and photometric observations of the cluster and
distributed populations. Constraints on the star formation history of 
molecular clouds can be obtained though by using the star count results 
from the previous section and examining the relative contribution of clusters 
and distributed populations to the total stellar population. The ages of 
several clusters in nearby molecular clouds have been inferred by constructing 
HR diagrams and comparing the stellar effective temperatures and luminosities 
with pre-main-sequence evolutionary tracks, including 
$\rho$ Ophiuchus \citep{Greene95,LR99}, \IC348 \citep{Luhman98,Herbig98}, 
the Orion Nebula Cluster \citep{Lynne97}, \NGC2024 \citep{Meyer96}, 
L1641~N \citep{Hodapp93,Allen96}, and L1641 South \citep{Allen96}. A literal 
interpretation of the HR diagrams indicate that most of the stars in these
clusters have formed within the last 1-2~Myr, although some stars may be 
as old as 10~Myr. By comparison, \co\ observations of OB associations 
\citep{Bash77,Leisawitz90} and the observed fraction of molecular clouds with
current star formation suggest that the cloud lifetime is between 10~Myr and 
100~Myr \citep{Elmegreen91}. Even longer lifetimes though are required in 
coagulation models for cloud formation \citep{Kwan83}. Relative to the 
expected lifetime of 
molecular clouds then, embedded clusters represent a recent ``burst'' of star 
formation. A distributed stellar population cannot be interpreted 
unambiguously, as it can represent either older clusters that have since 
dispersed or stars of all ages that formed in isolation throughout the cloud. 
The relative number of the total stellar population (cluster plus distributed) 
currently contained in clusters then places a lower limit on the fraction of 
stars that have formed in the recent history of a molecular cloud.

\subsection{Sensitivity}
\label{dis:sensitivity}

Before examining the implications of the star count results for the star 
formation history of the molecular clouds studied here, it is important to 
recognize the sensitivity of the \MASS\ observations to stars of various 
masses and ages. These dependencies are examined in Figure~\ref{sensitivity}, 
which shows \KB\ band iso-magnitude curves for stars as a function of stellar 
age ($10^5 - 10^8$ years) and stellar mass ($0.08-3.0$\msun) at the distance 
of Perseus, Orion~A/B, and MonR2. The magnitudes were computed using the 
pre-main-sequence evolutionary tracks from D'Antona \& Mazzitelli (1997,98)
assuming zero extinction and no near-infrared excess as outlined in
\citet{Lynne00}. Only the \KB\ band sensitivity limits are considered 
here since, as shown in Section~\ref{distributed}, the \KB\ band star
counts provide the tightest constraint on the surface density of the 
distribution population. Figure~\ref{sensitivity} shows, for example, that at 
the distance of the Perseus molecular cloud, 0.08\msun\ stars are brighter 
than the adopted \KB\ band magnitude threshold for ages \aboutless 10~Myr. 
(Young brown dwarfs with masses $<$ 0.08\msun, if present, are also 
detected at these distances for young ages. The shape of the initial mass 
function for these objects remains uncertain at this time, however, and these 
low masses are not considered here.) At older ages, low mass pre-main-sequence 
stars become systematically fainter than the sensitivity limits of the survey 
and would no longer be detectable. Assuming that clouds have ages \aboutless 
100~Myr, Figure~\ref{sensitivity} shows that all stars more massive than 
\about 0.3\msun\ on the surface of the Perseus molecular clouds will be 
detectable, with corresponding limits of \about 0.5\msun\ for Orion~A and 
Orion~B and \about 0.7\msun\ for MonR2.

The fraction of the total stellar population that is detectable with \MASS\ 
depends on the mass and age distribution of the constituent stars. These 
distributions are crudely known for only a handful of clusters and are 
completely unknown for the distributed population. Therefore, we have assumed 
that the stars have been forming at a constant rate in time throughout the
cloud with a Miller-Scalo Initial Mass Function \citep{MS79} between 0.08 and 
10\msun. The magnitudes for this model stellar population were calculated 
using the pre-main-sequence tracks up to 3\msun\ as described above. Stars 
more massive than 3\msun\ are not included in the \citet{DM97} models, but 
since their pre-main-sequence phase lasts for only \aboutless 2~Myr 
\citep{PS93}, the magnitudes of these stars as they appear on the zero age
main sequence were adopted for all ages. The average visual extinction 
computed from the \thco\ maps (3.0\M, 3.2\M, and 2.6\M\ for Perseus, 
Orion~A/B, and MonR2 respectively) were applied to the computed magnitudes. 

Figure~\ref{fraction} shows the fraction of the model stellar population that
has apparent \KB\ magnitudes between 6.0\M\ and 14.3\M\ as a function of the
time. While the model is intended mainly to describe a 
distributed population forming over an extended period of time, the results 
can be applied to clusters as well. This figure indicates that for clusters, 
which have typical ages \aboutless 2~Myr as discussed above, \aboutmore 80\% 
of the cluster members with stellar masses $\ge$ 0.08\msun\ are detectable 
with \MASS\ for each of the molecular 
clouds studied here. (Note, however, that clusters often have high extinction 
assumed for this model.) On the other hand, for a distributed population that 
has been forming stars for 50~Myr, 67\% of the stars would be detectable in 
Perseus, 45\% in Orion~A and Orion~B, and only 26\% in MonR2. This model 
demonstrates that if the distributed population has been forming for over tens 
of million of years, an appreciable fraction of the stars would no longer be 
detectable within the adopted magnitude thresholds.

\subsection{Star Formation Efficiency}

Given the above calculations, the observed surface density can be used to 
place constraints on the time averaged star formation efficiency in molecular 
clouds, where the star formation efficiency is defined as the fraction of the 
cloud mass that has been converted into stars. The predicted surface density of
stars can be computed using the cloud masses and angular areas summarized in
Table~\ref{tbl:clouds}, and assuming that the stars have been forming at a 
constant rate in time with a Miller-Scalo Initial Mass Function between 
0.08\msun\ and 10\msun. The solid curves in Figure~\ref{sfe} show the 
predicted \KB\ surface density for stars with magnitudes of 6.0\M\ $\le m(K_s) 
\le 14.3$\M\ for the nominal extinction model assuming star formation 
efficiencies of 1\%, 2\%, 5\%, and 8\%. The dashed lines in Figure~\ref{sfe} 
show the observed stellar surface density for the distributed population only. 
Similarly, the dotted lines show the surface density of the clusters only, 
averaged over the entire cloud area in order to determine how efficient mass 
is being converted into clusters on a global scale. The star formation 
efficiencies in the dense cores that the clusters formed in will naturally be 
higher. As with Figure~\ref{fraction}, the older ages are intended to apply 
only to the distributed population and not the clusters. Figure~\ref{sfe} 
indicates that the star formation efficiency implied by the distributed 
population alone is $<$1\% for Perseus, 1-5\% for Orion~A, 0\% for Orion~B, 
and 1-4\% for MonR2, where the range is due to the uncertain ages of the 
distributed population. The star formation efficiency implied by the clusters 
only is \about 2\% in Perseus, 4\% in Orion~A, 1\% in Orion~B, and 1\% in 
MonR2. The total star formation efficiency implied by the sum of the cluster 
and distribution populations then ranges from a low of \about 1\% in Orion~B 
to a high of 9\% in Orion~A if the distribution population is as old as 
100~Myr. 

\subsection{The Fraction of Stars in Clusters}

Since the observed ages of clusters is \aboutless 2~Myr, the fraction of the 
total stellar population contained in clusters alone reflects the relative 
contributions of recent star formation activity compared to isolated star 
forming regions and older stars that constitute the distributed population. 
The fraction of the detected stellar population contained in clusters using 
the nominal extinction model is 80\%, 73\%, 100\%, and 56\% for Perseus, 
Orion~A, Orion~B, and MonR2 respectively, where the contributions to the 
stellar population from 
previously known clusters not contained in the \MASS\ Second Incremental 
Release have been included (see Section~\ref{clusters}). The cluster fraction 
may be as high as 100\%, 87\%, 100\%, and 74\% for these clouds if the low 
extinction model is adopted, and as low as 34\%, 59\%, 39\%, and 41\% with the 
high extinction model. As emphasized above, however, an appreciable fraction
of the  distributed population may not have been detected if the stellar 
population has been forming for more than several million years. Thus the 
observed fraction of stars in clusters may underestimate the contribution from 
the distributed population. The number of stars in the distributed population 
that are too faint to detect with the present observations was estimated in 
Figure~\ref{fraction} assuming a constant star formation rate and a 
Miller-Scalo IMF. Applying these assumptions and model calculations, 
Figure~\ref{clusterfraction} shows the implied fraction of the total stellar 
population contained in clusters as a function of age for the distributed 
population. The solid curve shows the fraction of stars in clusters for the
nominal extinction model, and the dotted curves represent the cluster fraction
for the low and high extinction models. After taking into account the ages of 
the distributed population, the fraction of stars contained in clusters under 
the nominal extinction model is \aboutmore 50\% in Perseus, Orion~A, and 
Orion~B for ages less than 100~Myr, and \aboutmore 25\% for MonR2. Even for 
the high extinction model and an old age (100~Myr), the fraction of stars in 
clusters is still between 17-41\% for the four clouds. Thus in each of these 
molecular clouds, a substantial fraction of the total stellar population, if
not the majority, is contained in young stellar clusters.

\subsection{Discussion}

The substantial fraction of stars found in clusters regardless of the age of 
the distributed population is surprising given the canonical assumed lifetimes
of molecular clouds. As discussed above, embedded, dense clusters typically 
have ages \aboutless 2~Myr. The fact that no substantially older clusters are 
found within molecular clouds indicates that these older clusters have 
dispersed into the molecular cloud, have destroyed the surrounding molecular
gas and no longer appear associated with a cloud, or these clouds are just now
forming their first clusters. If we assume a constant star formation rate
and that clusters simply disperse after a few million years, then the fraction
of stars contained in clusters should decrease as $t^{-1}$, where $t$ is the
age of the molecular cloud. With these assumptions and assuming a cloud 
lifetime of 50~Myr just as an example, the fraction of stars in clusters 
should be \about 4\%. By contrast, the observed fraction of stars in clusters 
for the clouds analyzed here is substantially higher as discussed above, and 
would imply a cloud lifetime of \aboutless 7~Myr under the constant star 
formation rate assumption even given the uncertainties in the field star 
subtraction. Thus in each of the clouds,
the distributed population contains fewer stars than expected if star formation
has been occurring in these regions at a constant rate for tens of millions of 
years. This situation is reminiscent of the post-T Tauri star problem in the 
Taurus molecular cloud, where most stars within the confines of the molecular 
gas have ages \aboutless 3~Myr and few stars have been identified that are 
older than 10~Myr \citep{Herbig78}. Various hypotheses have been proposed to 
explain the lack of older stars in Taurus and other nearby molecular clouds, 
including the kinematic dispersal of older stars from the molecular cloud 
boundaries, imposing a star formation rate that changes with time, and 
postulating that molecular clouds are inherently short lived. We briefly 
discuss these possibilities as they pertain to the Perseus, Orion~A, Orion~B, 
and MonR2 molecular clouds.

The small number of old stars that have been identified in nearby molecular 
clouds has led to the suggestion that a substantial fraction of the stellar 
population has dispersed from the molecular gas and that the current stellar
census is incomplete \citep{Feigelson96}. Assuming a one dimensional velocity 
of 1~\kms, a star could drift as far as 10~pc (1.2\arcdeg\ at the distance of 
Orion~A and Orion~B) in 10~Myr. However, molecular clouds are likely 
gravitationally bound structures, and presumably the kinematics of the newly 
formed stars reflect that of the molecular gas. Since the star formation 
efficiency implied by the star counts is \aboutless 9\%, the dynamical 
evolution of the embedded stars will be dominated by the gas as long as the 
cloud remains intact. Therefore, many, if not most, of the stars must remain 
within the cloud. Nevertheless, the size of any dispersed stellar population 
can by gauged from H$\alpha$ and X-ray surveys of regions surrounding the 
cloud. This has been in fact done for the Orion~A and Orion~B molecular 
clouds, but the results remain inconclusive. Both H$\alpha$ \citep{Nakano95} 
and X-ray \citep{Sterzik95} surveys of this region find a source density of 
\about 1~star~deg$^{-2}$ in regions surrounding the Orion molecular clouds and 
OB association. Up to one-third of the X-ray sources may be unrelated 
foreground stars \citep{Alcala00,Alcala96,Alcala98} perhaps related to Gould's 
belt \citep{G98}, but in any event, the surface density of sources is too low 
to contribute significantly to the total population in the Orion~A and Orion~B
molecular clouds relative to the clusters. H$\alpha$ emission is likely not 
sensitive to the majority of the older stars likely to constitute a 
old dispersed population, though, and at the distance of the Orion clouds, the 
ROSAT All Sky Survey is able to detect only to the more luminous X-ray T Tauri 
stars \citep{Neu95,Alcala00}. In more nearby clouds, however, follow up 
studies of the widely distributed ROSAT X-ray sources indicate that many these 
objects likely do not represent a dispersed population, but rather solar-type 
zero age main sequence stars or older pre-main-sequence stars unrelated to the 
existing molecular clouds (Magazz\`u \etal 1997; see also Brice\~no \etal 
1997, Favata, Micela, \& Sciortino 1997). Further, a large scale X-ray and 
objective-prism survey of Taurus failed to find any new T Tauri stars older 
than 5~Myr \citep{Briceno99}. 

If most of the stars that have formed over the lifetime of the Perseus,
Orion~A, Orion~B, and MonR2 molecular clouds are still associated with the
molecular gas, then the high fraction of the total stellar population 
contained in clusters implies that either the star formation rate is currently
higher than it has been in the past \citep{Palla97,PS00}, or that molecular 
clouds have relatively young ages \citep{BP99,Elmegreen00}. \citet{PS00} 
re-analyzed published HR diagrams for a number of nearby star forming regions 
with a single set of pre-main-sequence models and argued that star formation 
over entire molecular clouds (e.g. Taurus and Chamaeleon~I) and individual 
clusters (e.g. the Orion Nebula Cluster) started at a low rate \aboutmore 
10~Myr ago and has increased dramatically within the past 1-3~Myr. In their 
scenario, molecular clouds remain relatively dormant for much of their cloud 
lifetime since individual dense cores are supported against gravitational 
collapse by magnetic fields, and the time scale to dissipate the magnetic 
support is on the order of \about 10~Myr (see also Palla \& Galli~1997). 
\citet{BP99}, on the other hand, suggested that the lack of old stars in 
Taurus this is a consequence of the molecular filaments that make up the cloud
having formed only within the past few million years in turbulent flows in the
interstellar medium. In this scenario, Taurus is an intrinsically young
cloud that has not had time to form old stars (see also Elmegreen~2000). 

The difference between the two pictures for the evolution of molecular clouds
has important consequences for molecular cloud evolution. In the 
\citet{Palla97} scenario, molecular clouds are in a quasi-static state for 
most of their lifetime, while the \citet{BP99} model suggest clouds form and 
evolve dynamically in the interstellar medium. If the \citet{PS00} model is 
correct, most molecular clouds should not contain a substantial stellar 
population since they are relatively dormant for most of their cloud 
lifetimes. The difficulty with this scenario is that it would imply that we 
are observing a number of molecular clouds in the solar neighborhood just
as they are forming a substantial fraction of their total stellar population 
in young clusters. This would include not only the 4 molecular clouds studied 
here, but also Chamaeleon~I, Lupus, and Taurus. This is not necessarily an 
observational bias in that these 
clouds are frequently studied because they have active star formation, as star
formation and stellar clusters appear to be ubiquitous in molecular clouds.
More than half of the clouds within 200~pc of the sun are forming an 
appreciable number of T Tauri stars (Feigelson~1996 and references therein),
and most Giant Molecular Clouds within 3~kpc of the sun are forming OB stars 
and presumably an accompanied clusters of lower mass stars \citep{Blitz91}. In 
this respect, imposing a short lifetime for molecular clouds would 
qualitatively account for the lack of a substantial older stellar population 
in molecular clouds without the need to impose a higher than average star 
formation rate at the current time \citep{BP99,Elmegreen00}. A more systematic 
survey of a volume limited sample of molecular clouds seems warranted though 
to better establish the frequency of clusters within molecular clouds as a 
more comprehensive observational test of these two pictures of cloud 
evolution.

\section{Summary}
\label{summary}

We investigate the spatial distribution of young stars associated with the 
Perseus, Orion~A, Orion~B, and MonR2 molecular clouds using \JB, \HB, and 
\KB\ star counts from the \MASS\ Second Incremental Release. The stellar
population associated within these clouds is determined statistically by 
subtracting a semi-empirical model for the field star contamination from the
observed stellar surface density maps. The field star model is constructed 
using \MASS\ observations of the regions surrounding the clouds to measure the 
total field star surface density, published \thcoj\ maps to determine the cloud 
extinction, and the \citet{Wainscoat92} Galactic star count model to estimate 
the fraction of field stars that are background to the clouds. The stellar 
population is categorized into compact stellar clusters and a distributed
stellar population. Individual stellar clusters are identified as peaks in the 
field star subtracted, stellar surface density maps, and the distributed 
stellar population is defined as any excess star counts outside the cluster 
boundaries but within the molecular cloud as traced by \thcoj\ emission.

The number of clusters within the studied molecular clouds, including clusters 
identified in the literature in regions not yet observed with \MASS, 
varies from 2 in Perseus to 7 in Orion~A. The total number of stars contained 
within the clusters is $>$393 in Perseus, $>$1992 in Orion~A, 606 in Orion~B, 
and 543 in MonR2. More than half of the cluster members in each cloud is 
contained in just a single rich cluster. A distributed stellar population is 
detected in the Orion~A and MonR2 molecular clouds within the uncertainties
of the fieldstar subtraction at a \KB\ band surface density between 
0.013\sqamin\ and 0.084\sqamin. Model calculations suggest, however, that the 
surface density of the distributed population in the clouds may be 
underestimated by as much as a factor of three if the stars have been forming 
with a Miller-Scalo Initial Mass Function between 0.08\msun\ and 10\msun\ at a 
constant star formation rate for up to 100~Myr. After taking into 
consideration the possible evolutionary status of the distributed population, 
the star formation efficiencies implied by the sum of cluster and distributed 
populations varies between \about 1-9\% for the four clouds. The percentage of 
the total observed stellar population contained in clusters alone is 80\%, 
73\%, 100\%, and 56\% for Perseus, Orion~A, Orion~B, and MonR2 respectively 
for the nominal extinction model, but may be up to a factor of two lower 
depending again on the age of the distributed population. Nonetheless, the 
high fraction of stars currently contained in clusters is surprising given 
that embedded clusters typically have ages (\aboutless 2~Myr) that are 
substantially younger than that often assumed for molecular clouds 
(10-100~Myr). These results indicate that either each of these molecular 
clouds have been forming stars at a higher rate in the last couple of million 
years than in the prior history, or that the ages of molecular clouds are 
younger than generally assumed.

\acknowledgements

JMC would like to thank Lynne Hillenbrand for numerous discussions and 
comments on this work. He is also grateful to John Bally for providing the 
\thco\ molecular line maps. This publication makes use of data products from 
the Two Micron All Sky Survey, which is a joint project of the University of 
Massachusetts and the Infrared Processing and Analysis Center, funded by the 
National Aeronautics and Space Administration and the National Science 
Foundation. JMC acknowledges support Long Term Space Astrophysics Grant 
NAG5-8217 and the Owens Valley Radio Observatory, which is supported by
the National Science Foundation through NSF grant AST 9981546.

\clearpage

\begin{figure}
\caption{
  A \KB\ band stellar surface density map for point sources in the \MASS\ 
  Second Incremental Release for the area between Galactic longitudes of
  130\arcdeg\ and 250\arcdeg\ and Galactic latitudes of $-40$\arcdeg\ and 
  $+40$\arcdeg. The map was produced by binning stars with magnitudes
  $6.0^{\rm m} \le m(K_s) \le 14.3^{\rm m}$ in $5'\times5'$ pixels. The map
  is presented in the Hammer-Aitoff projection with a square-root image 
  stretch. Darker gray scales denote higher stellar surface densities, with 
  the surface densities ranging from \about 2\sqamin\ in the Galactic Plane to 
  \about 0.2\sqamin\ at higher Galactic latitudes. The white regions represent 
  tiles not included in the Second Incremental Release, or in less than 1\% of 
  the instances, tiles not meeting the sensitivity criteria adopted for this 
  paper (see text). The location of several well known molecular clouds are 
  indicated.
  \label{ksqrt}
}
\end{figure}
\clearpage

\begin{figure}
\caption{
  Image of the average \JK\ stellar color over the same region shown in 
  Figure~\ref{ksqrt}. The darker gray scales represent larger \JK\ colors and 
  indicate regions where background field stars and embedded young stars have
  been reddened by dust in molecular clouds. The Taurus, Perseus, Orion~A, 
  Orion~B, and MonR2 molecular clouds are clearly visible, as well as numerous
  molecular clouds within the Galactic Plane.
  \label{jk}
}
\end{figure}
\clearpage

\begin{figure}
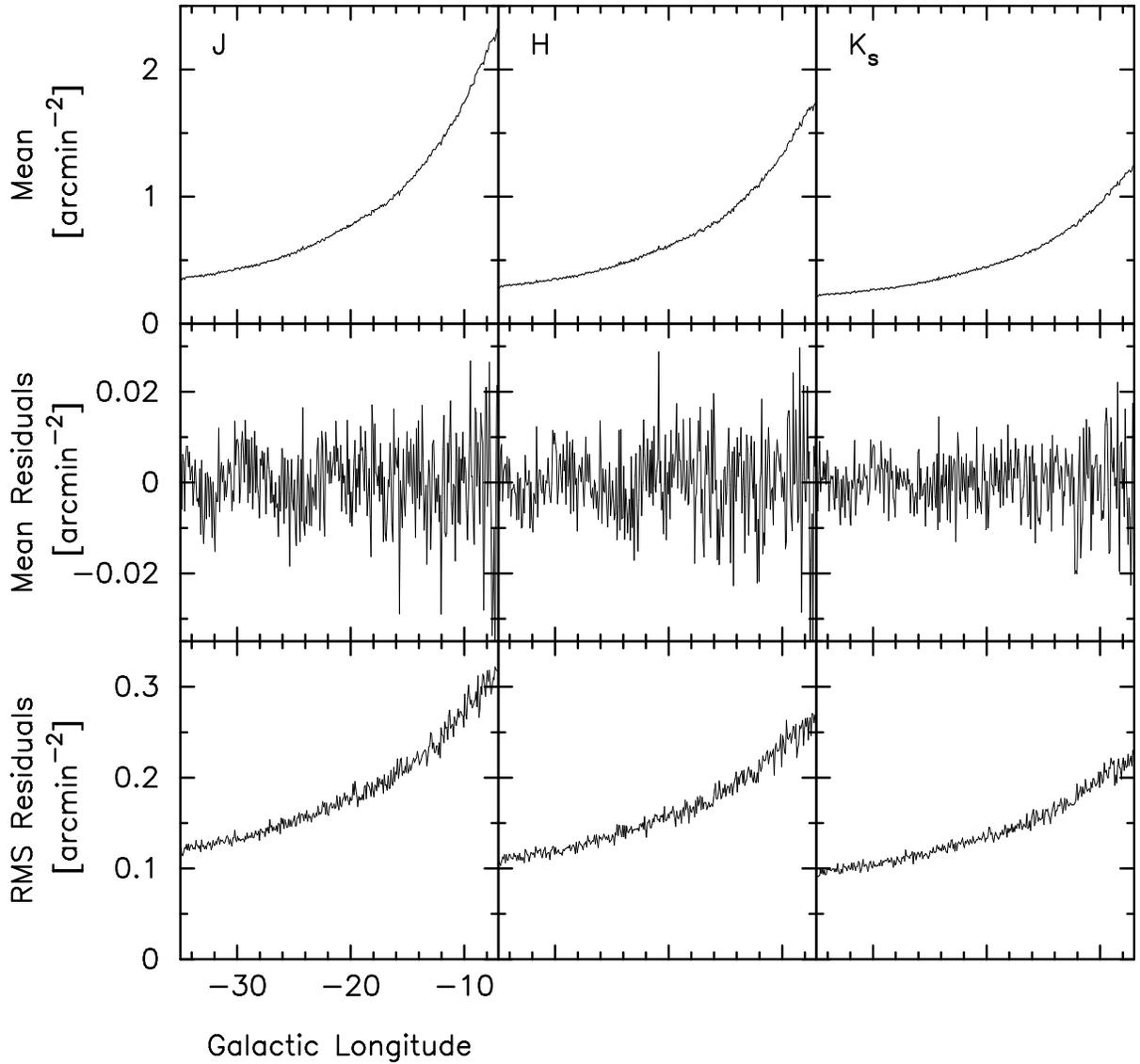

\insertplot{figure03.ps}{5.0}{4}{-2.0}{2.7}{0.85}{1}
\vspace{2.0truein}
\caption{
  The observed mean stellar surface density as a function of Galactic latitude
  (top panels), the mean residuals after subtracting a polynomial fit to the 
  observed star counts (middle panels), and the RMS of the residuals (bottom 
  panels) at \JB, \HB, and \KB\ band. Only regions between Galactic longitudes 
  of 130\arcdeg\ and 250\arcdeg\ and latitudes of $-35$\arcdeg\ and 
  $-7$\arcdeg\ that did not contain a stellar cluster or a molecular cloud 
  are shown here and included in the polynomial fit. The fit was performed as 
  a function of both Galactic longitude and latitude, but are averaged over 
  longitude for presentation in this figure. For reference, the Perseus, 
  Orion~A, Orion~B, and MonR2 molecular clouds analyzed in this paper are 
  located between latitudes of -20\arcdeg\ and -10\arcdeg.
  \label{fit}
}
\end{figure}
\clearpage

\begin{figure}
\vspace{-1.3truein}
\caption{
  {\it Upper left:} 
    An adaptive kernel, \KB\ band stellar surface density map of the Perseus 
    molecular cloud for stars with magnitudes of 6.0\M\ $\le m(K_s) \le 
    14.3$\M.
  {\it Upper right:} 
    The \IRAS\ 60\micron\ image displayed in a logarithmic stretch. 
  {\it Lower left:} 
    An image of the average \JK\ color for stars observed by \MASS.
  {\it Lower right:} 
    A map of the integrated \thcoj\ intensity map \citep{Padoan99}.
  In each panel, darker halftones represent the higher intensities.
  In the \KB\ density map and the average \JK\ color image, the white strips 
  are tiles not included in the \MASS\ Second Incremental Release, and the 
  white ``crosses'' are regions around bright stars that were masked out 
  when generating the Second Incremental Release Point Source Catalog. The 
  labels indicate the location of either prominent star forming regions or 
  stellar clusters identified in the \KB\ surface density image (see 
  Table~\ref{tbl:clusters}).
  \label{perseus_kernel}
}
\end{figure}
\clearpage

\begin{figure}
\caption{
  Same as in Figure~\ref{perseus_kernel}, except for the Orion~A (L1641)
  molecular cloud. The \thco\ image is from \citet{Bally87}.
  \label{oriona_kernel}
}
\end{figure}
\clearpage

\begin{figure}
\caption{
  Same as in Figure~\ref{perseus_kernel}, except for the Orion~B (L1630)
  molecular cloud. The \thco\ image is from \citet{Miesch94}.
  \label{orionb_kernel}
}
\end{figure}
\clearpage

\begin{figure}
\caption{
  Same as in Figure~\ref{perseus_kernel}, except for the MonR2 molecular
  cloud. The \thco\ image is from \citet{Miesch94}. The sources labeled
  VDB are reflection nebula cataloged by \citet{VDB66}, and sources labeled 
  GGD are from the list of Herbig-Haro objects noted by \citet{GGD78}.
  \label{monr2_kernel}
}
\end{figure}
\clearpage

\begin{figure}
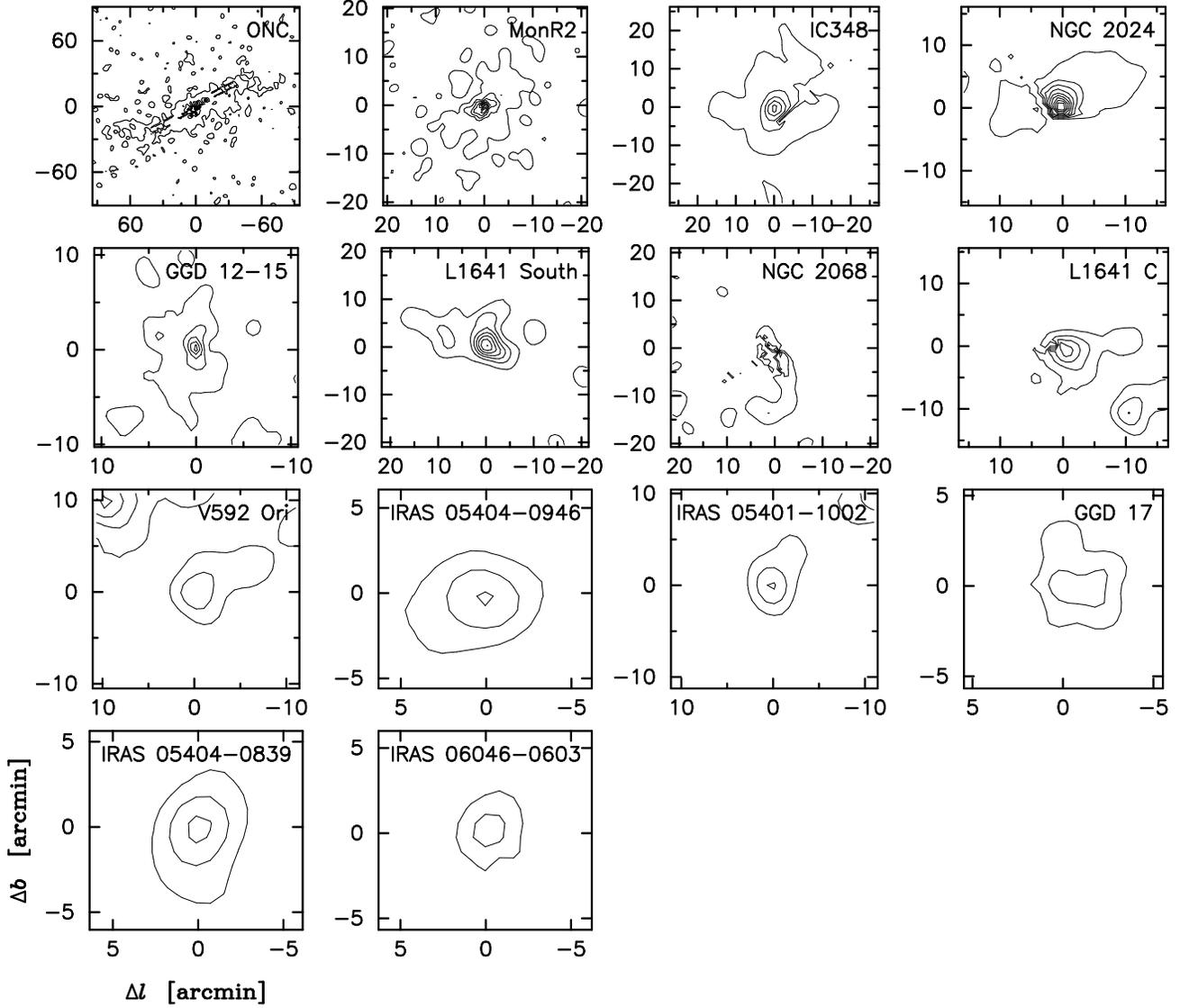

\insertplot{figure08.ps}{6.7}{4}{0.0}{3.0}{0.8}{1}
\caption{
  Contour maps of the \KB\ stellar surface density for each of the clusters
  summarized in Table~\ref{tbl:clusters} and labeled in 
  Figures~\ref{perseus_kernel}-\ref{monr2_kernel}. Each contour map is 
  centered on the cluster coordinates listed in Table~\ref{tbl:clusters}. The
  angular size of the contour maps varies depending on the dimensions of the 
  cluster. The contour levels in each map begin at 2$\sigma$ above the field 
  star stellar surface density, where $\sigma$ is the noise per pixel in the 
  field surface density map estimated assuming Poisson statistics. The contour 
  intervals are 10$\sigma$ for \IC348, ONC, \NGC2024, \GGD12-15, and MonR2, 
  and 3$\sigma$ for the remaining clusters. The approximate noise level in 
  field star surface density is $\sigma$ \about 0.1\sqamin\ for \IC348, \about 
  0.15\sqamin\ for clusters in Orion, and \about 0.3\sqamin\ for clusters in 
  MonR2. Note that part of the cluster area in \IC348, \LDN1641~C, \NGC2024, 
  and the ONC has been masked in the \MASS\ Second Incremental Release Point 
  Source Catalog to eliminate artifacts from bright stars. Also, the western 
  portion of the \NGC2068 cluster has not been imaged with \MASS\ at the time 
  of this study.
  \label{contours}
}
\end{figure}
\clearpage

\begin{figure}
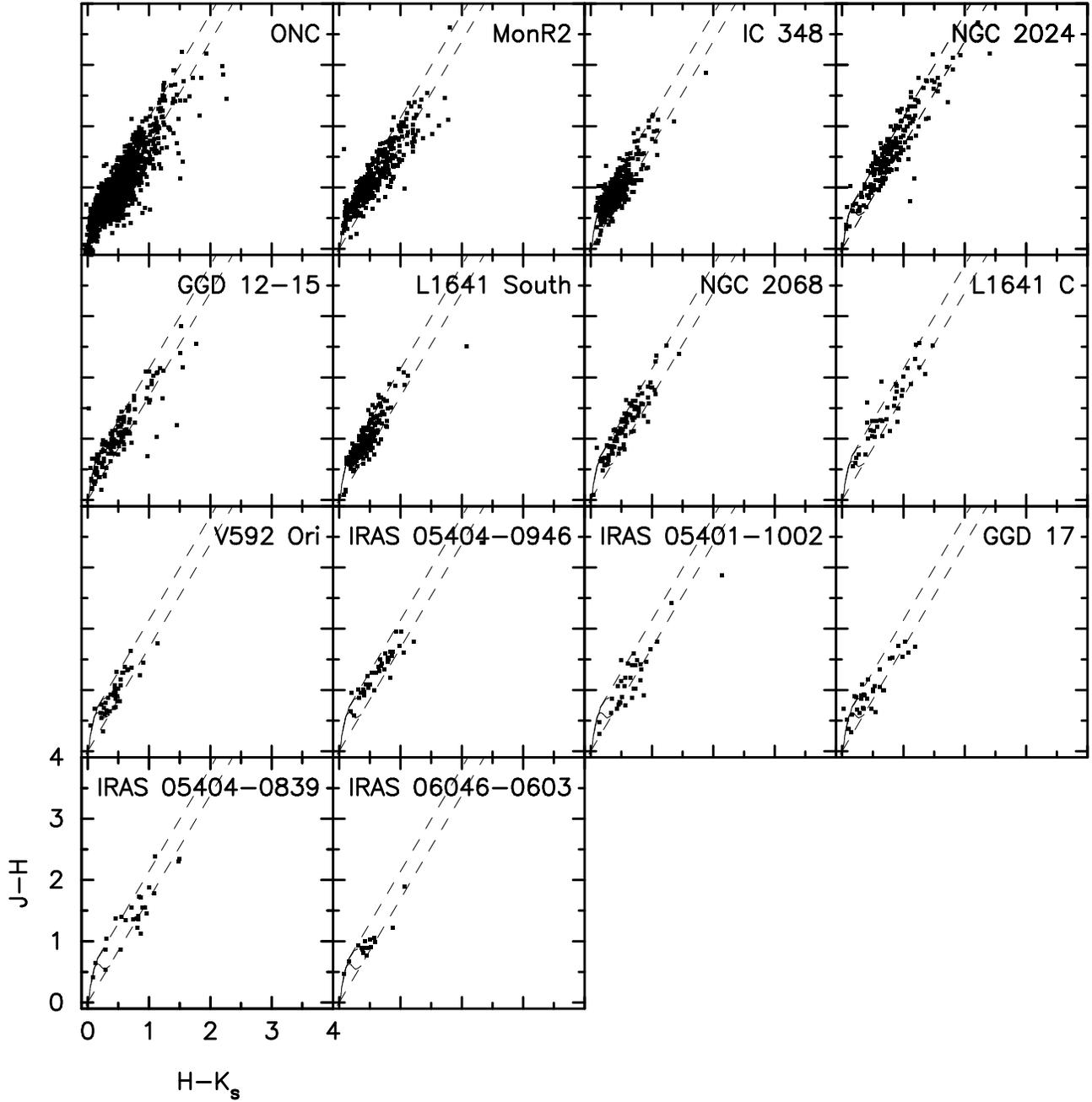

\insertplot{figure09.ps}{6.7}{4}{-0.3}{3.0}{1.0}{1}
\caption{
  $J-H$ vs $H-K_s$ color-color diagram for each cluster identified in the
  \KB\ band stellar density maps. Only stars that have magnitudes of 
  6.0\M\ $\le m(K_s) \le 14.3$\M\ and photometric uncertainties less than 
  0.1\M\ in all three bands are shown. The solid curves show the locus of
  unreddened main sequence and giant stars in the CIT system \citep{BB88},
  and the dashed lines show the reddening vectors from \citet{Cohen81}.
  Each cluster contains a number of red objects, supporting the notion
  that the clusters are indeed embedded within the molecular clouds. 
  \label{jhhk}
}
\end{figure}
\clearpage

\begin{figure}
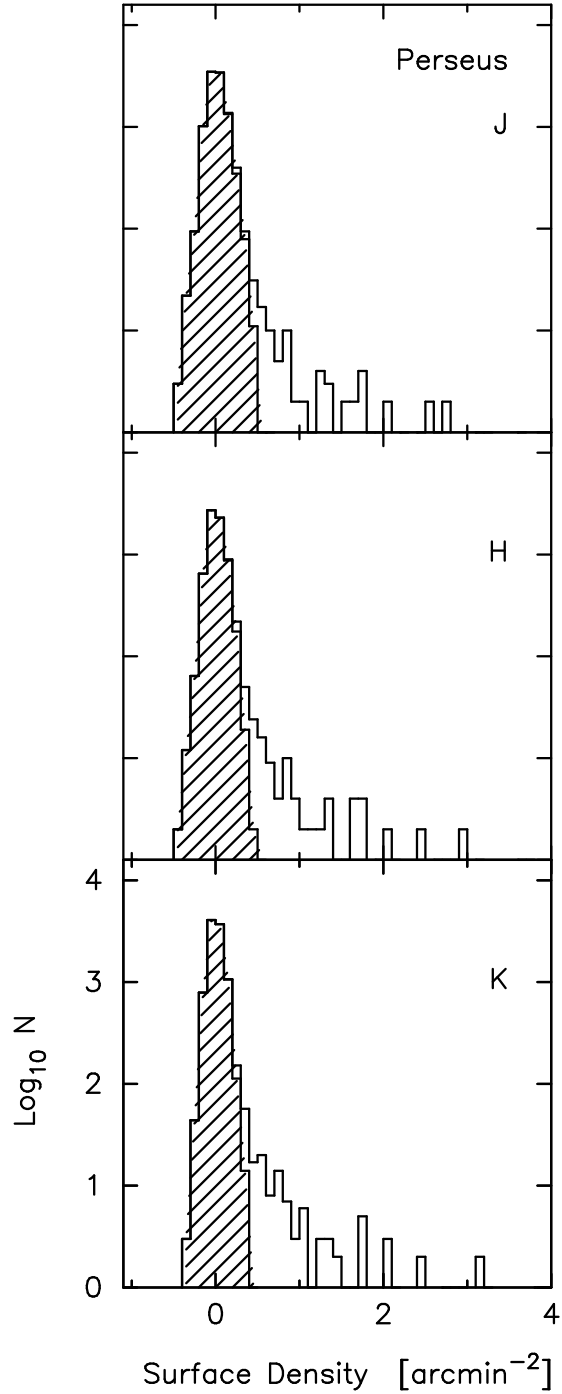

\insertplot{figure10.ps}{5.0}{4}{2.3}{4.5}{0.8}{0}
\vspace{3.0truein}
\caption{
  Histogram of the \JB, \HB, and \KB\ stellar surface densities observed
  toward the Perseus molecular cloud after subtracting the nominal field star 
  model. The open histograms are for all lines of sight toward the Perseus 
  molecular cloud as defined by \thcoj\ emission. The hatched regions 
  represent lines of sight outside the cluster boundaries but within the cloud
  area, and represent the surface density distribution of the distributed 
  population. The mean surface density of the distributed population as 
  inferred from the hatched histogram is summarized in 
  Table~\ref{tbl:distributed}.
  \label{perseus_hist}
}
\end{figure}
\clearpage

\begin{figure}
\insertplot{figure11.ps}{5.0}{4}{2.3}{4.5}{0.8}{0}
\vspace{3.0truein}
\caption{
  Same as Figure~\ref{perseus_hist}, except for the Orion A (L1641) molecular 
  cloud.
  \label{oriona_hist}
}
\end{figure}
\clearpage

\begin{figure}
\insertplot{figure12.ps}{5.0}{4}{2.3}{4.5}{0.8}{0}
\vspace{3.0truein}
\caption{
  Same as Figure~\ref{perseus_hist}, except for the Orion B (L1630) molecular 
  cloud.
  \label{orionb_hist}
}
\end{figure}
\clearpage

\begin{figure}
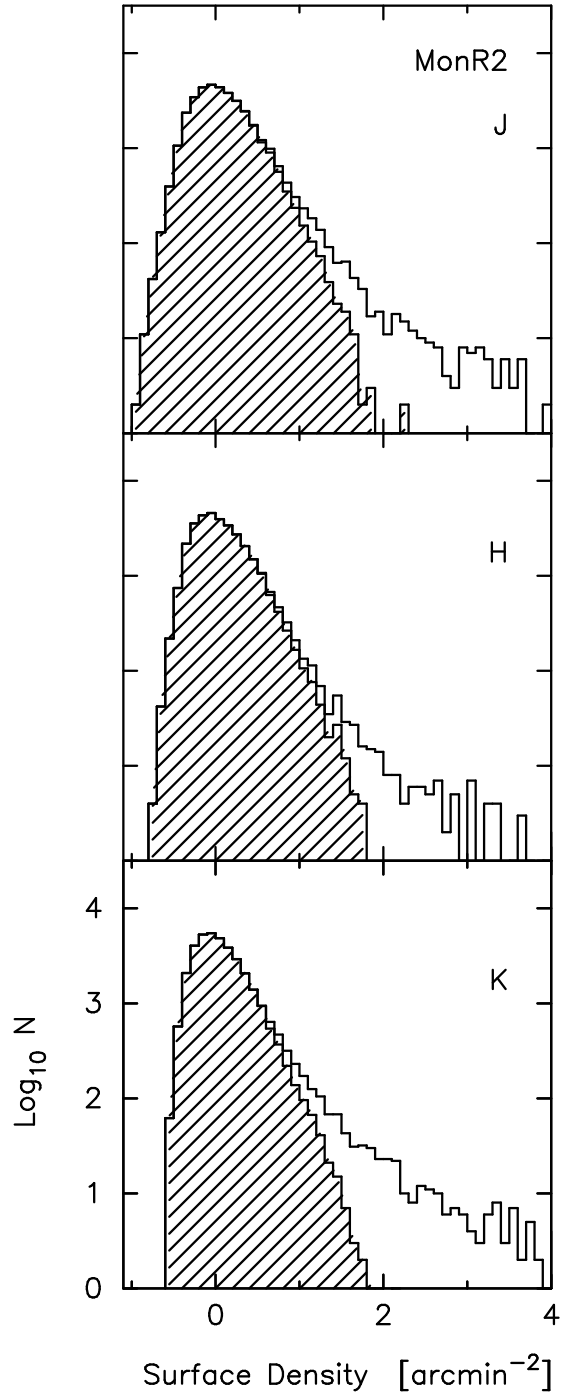

\insertplot{figure13.ps}{5.0}{4}{2.3}{4.5}{0.8}{0}
\vspace{3.0truein}
\caption{
  Same as Figure~\ref{perseus_hist}, except for the MonR2 molecular cloud.
  \label{monr2_hist}
}
\end{figure}
\clearpage

\begin{figure}
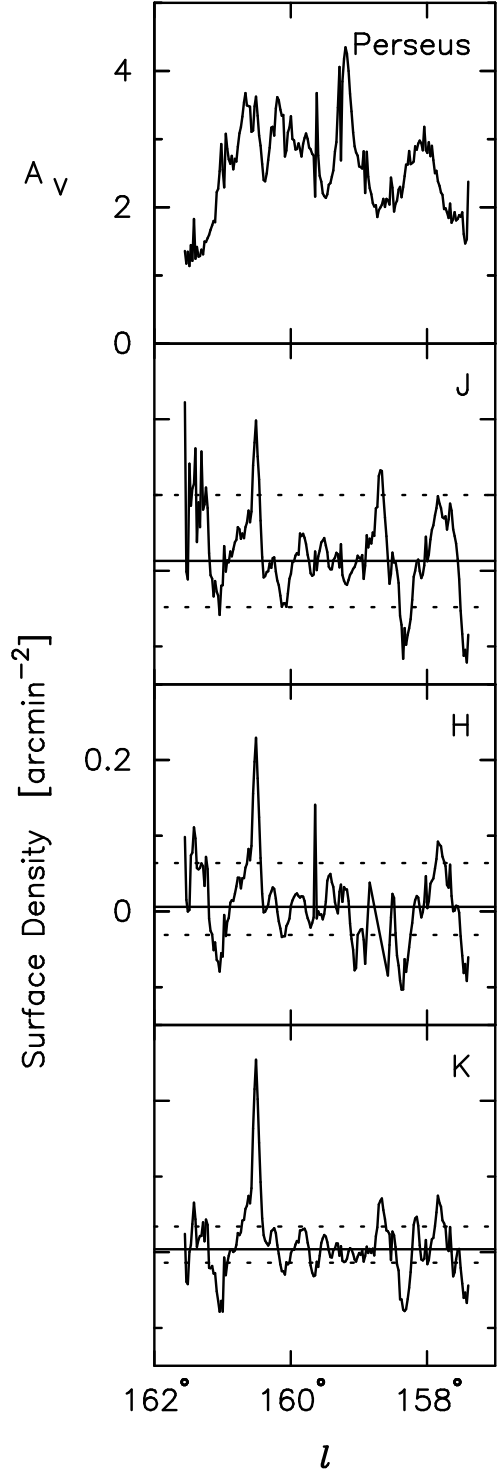

\insertplot{figure14.ps}{5.0}{4}{2.3}{4.5}{0.85}{0}
\vspace{3.0truein}
\caption{
  The visual extinction (top panel) and \JB, \HB, and \KB\ band stellar 
  surface density (bottom panels) as a function of Galactic longitude for the 
  Perseus molecular cloud. The field stars have been subtracted from the star
  counts using the nominal extinction model, and the results averaged over
  Galactic latitude. The solid horizontal line shows the average surface 
  density of the distributed population for the nominal extinction model, and 
  the dashed horizontal lines show the inferred surface density assuming the 
  low and high extinction models. This figure shows that the tightest 
  constraints on the surface density of the distributed population is 
  provided at \KB\ band where the field star subtraction is less sensitive
  to the assumed extinction model.
  \label{perseus_cuts}
}
\end{figure}
\clearpage

\begin{figure}
\insertplot{figure15.ps}{5.0}{4}{2.3}{4.5}{0.85}{0}
\vspace{3.2truein}
\caption{
  Same as in Figure~\ref{perseus_cuts}, but for the Orion A molecular cloud.
  \label{oriona_cuts}
}
\end{figure}
\clearpage

\begin{figure}
\insertplot{figure16.ps}{5.0}{4}{2.3}{4.5}{0.85}{0}
\vspace{3.2truein}
\caption{
  Same as in Figure~\ref{perseus_cuts}, but for the Orion B molecular cloud.
  \label{orionb_cuts}
}
\end{figure}
\clearpage

\begin{figure}
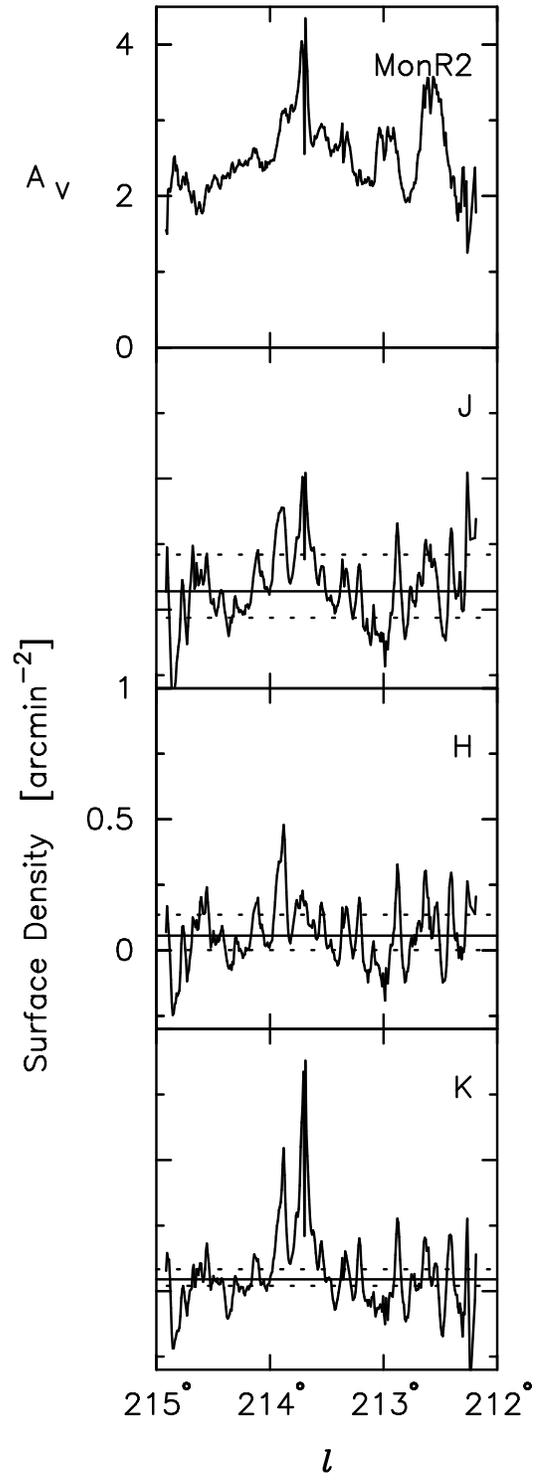

\insertplot{figure17.ps}{5.0}{4}{2.3}{4.5}{0.85}{0}
\vspace{3.2truein}
\caption{
  Same as in Figure~\ref{perseus_cuts}, but for the MonR2 molecular cloud.
  \label{monr2_cuts}
}
\end{figure}
\clearpage

\begin{figure}
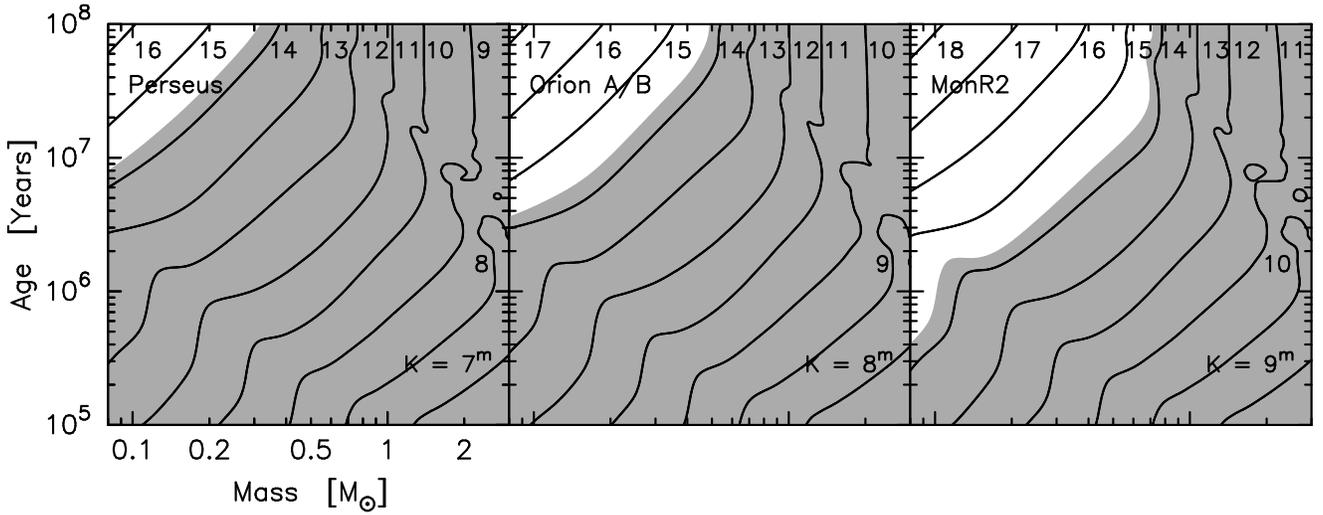

\insertplot{figure18.ps}{5.0}{4}{0.0}{3.0}{0.75}{1}
\vspace{-1.truein}
\caption{
   \KB\ band iso-magnitude contours as a function of stellar age and mass for
   stars at the distance of Perseus (320~pc), Orion A and Orion B (480~pc), 
   and MonR2 (830~pc). The magnitudes were computed using the D'Antona \& 
   Mazzitelli (1997, 1998) pre-main-sequence evolutionary tracks assuming 
   ${\rm A_V = 0^m}$ and that no near-infrared excess emission is present.
   The shaded area highlights the parameter space that is probed for the 
   adopted \KB\ magnitude thresholds (6.0\M\ $\le m(K_s) \le 14.3$\M).
  \label{sensitivity}
}
\end{figure}
\clearpage

\begin{figure}
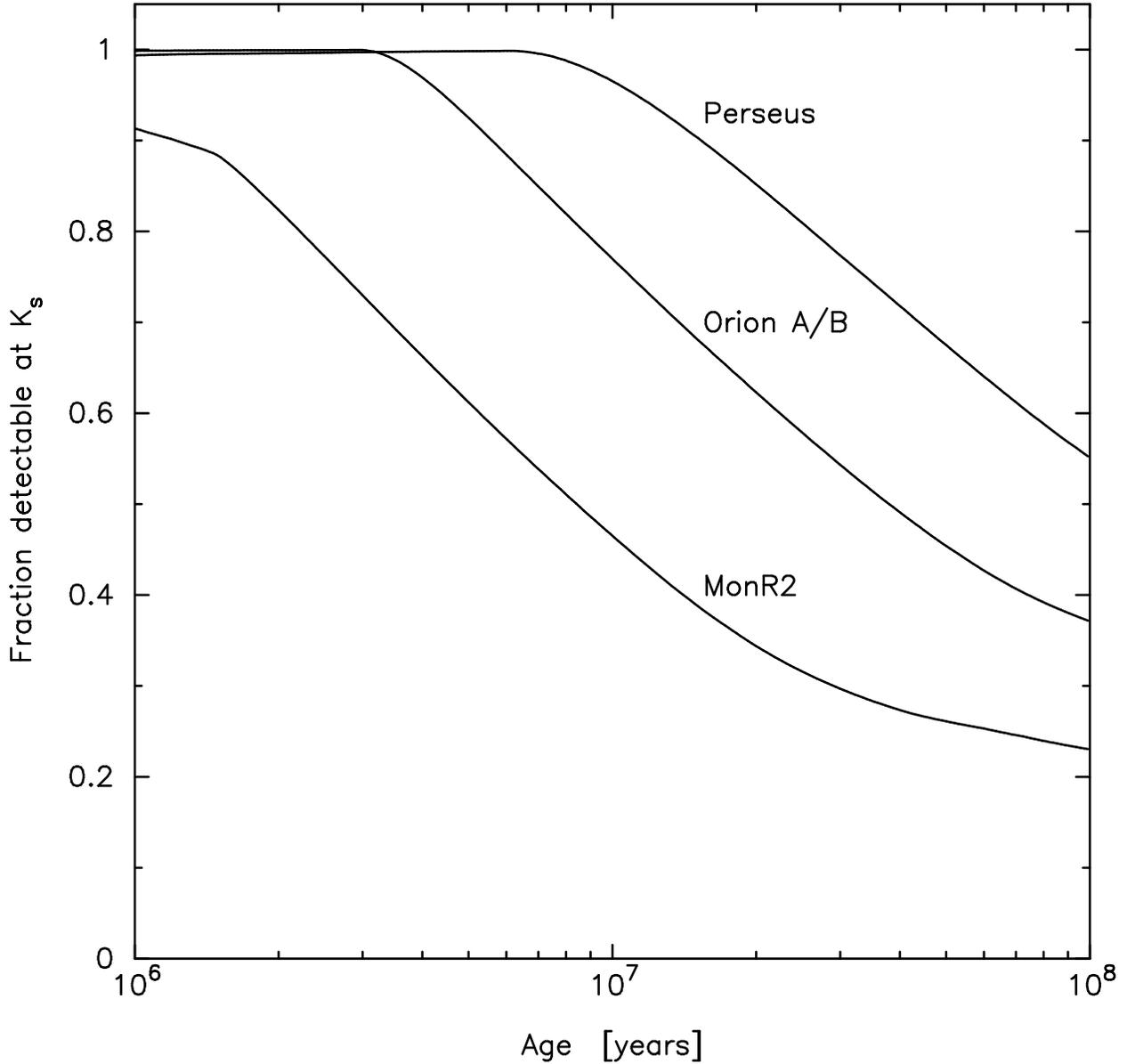

\insertplot{figure19.ps}{5.0}{4}{-2.0}{3.0}{0.9}{1}
\vspace{2.0truein}
\caption{
   The fraction of a model stellar population with magnitudes of 
   $6.0^{\rm m} \le m(K) \le 14.3^{\rm m}$ as a function of the molecular 
   cloud age for the Perseus, Orion~A/B, and MonR2 molecular clouds. The
   model assumes that stars have been forming at a constant at rate in time
   with a Miller-Scalo Initial Mass Function over the mass range of 
   0.08--10\msun, and that the visual extinction is the average value inferred 
   from the \thco\ maps ($A_V$ \about 3\M). The magnitudes were computed using 
   the D'Antona \& Mazzitelli (1997, 1998) pre-main-sequence evolutionary 
   tracks for objects with masses up to 3\msun, and assuming main sequence
   magnitudes for the more massive stars. The results indicate that the
   fraction of the model stellar population within the adopted \KB\ magnitude 
   thresholds for ages $<$ 100~Myr ranges from $>$ 67\% at the distance of 
   Perseus to $>$ 26\% at the distance of MonR2.
  \label{fraction}
}
\end{figure}
\clearpage

\begin{figure}
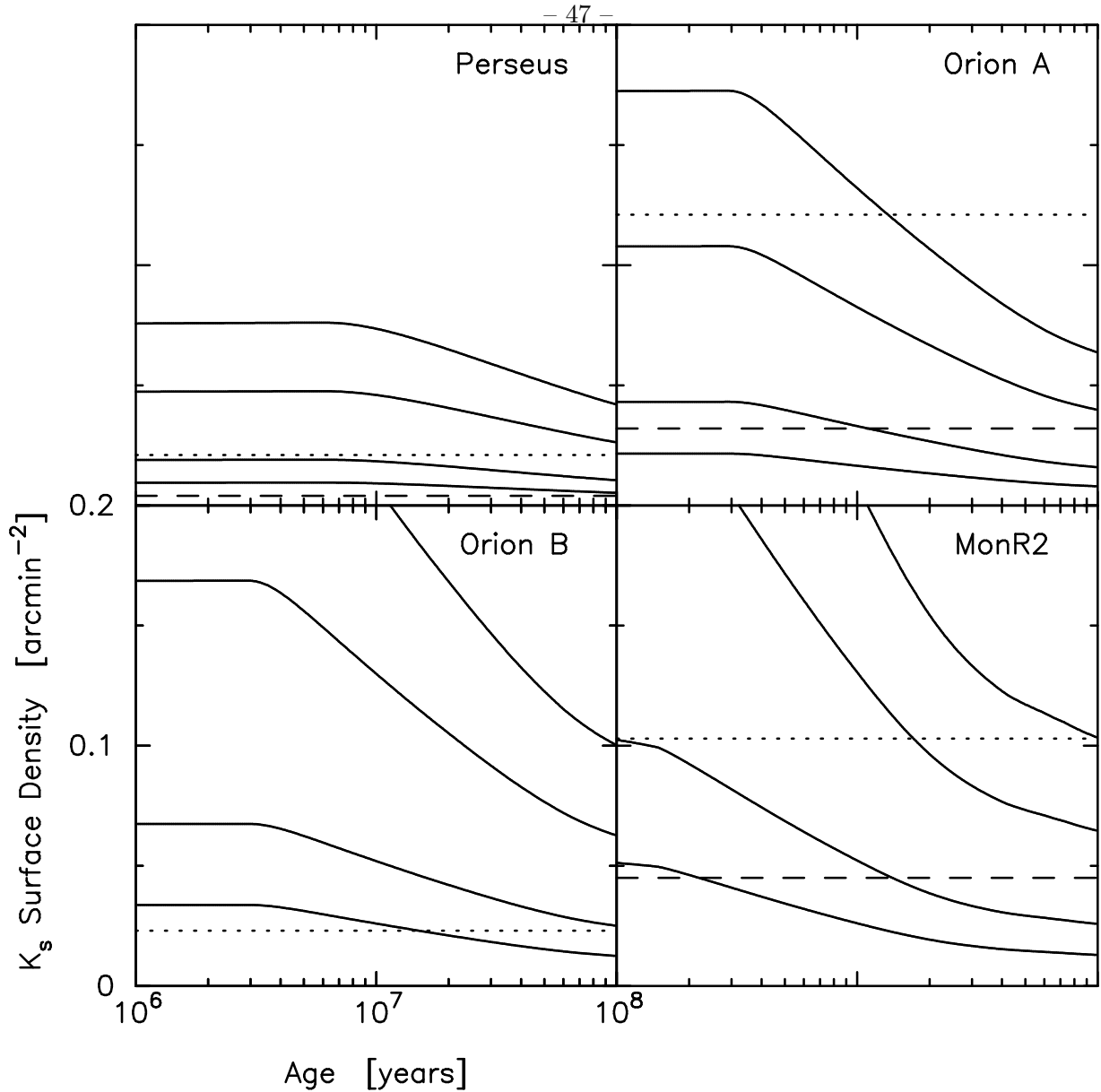

\insertplot{figure20.ps}{5.0}{4}{-1.5}{3.0}{0.9}{1}
\vspace{1.2truein}
\caption{
   The predicted surface density of stars with apparent magnitudes of
   $6.0^{\rm m} \le m(K) \le 14.3^{\rm m}$ as a function of the molecular 
   cloud age for the Perseus, Orion~A, Orion~B, and MonR2 molecular clouds for 
   the model calculations described in Figure~\ref{fraction}. The 
   solid curves are the predicted stellar surface density for an assumed 
   star formation efficiency of 1\% (bottom curves in each panel), 2\%, 5\%,
   and 8\% (top curves in each panel) using the cloud masses implied by 
   the nominal extinction model. The horizontal dashed line shows the inferred 
   surface density for the distributed population (if a positive value; see 
   Table~\ref{tbl:distributed}), and the horizontal dotted line shows surface 
   density for the clusters averaged over the entire cloud area. The 
   older ages are intended to apply only to the distributed population. The
   global star formation efficiency implied by the sum of the cluster and 
   distributed population ranges between \about 2 and 9\% for the four 
   clouds depending on the age assumed for the distributed population.
  \label{sfe}
}
\end{figure}
\clearpage

\begin{figure}
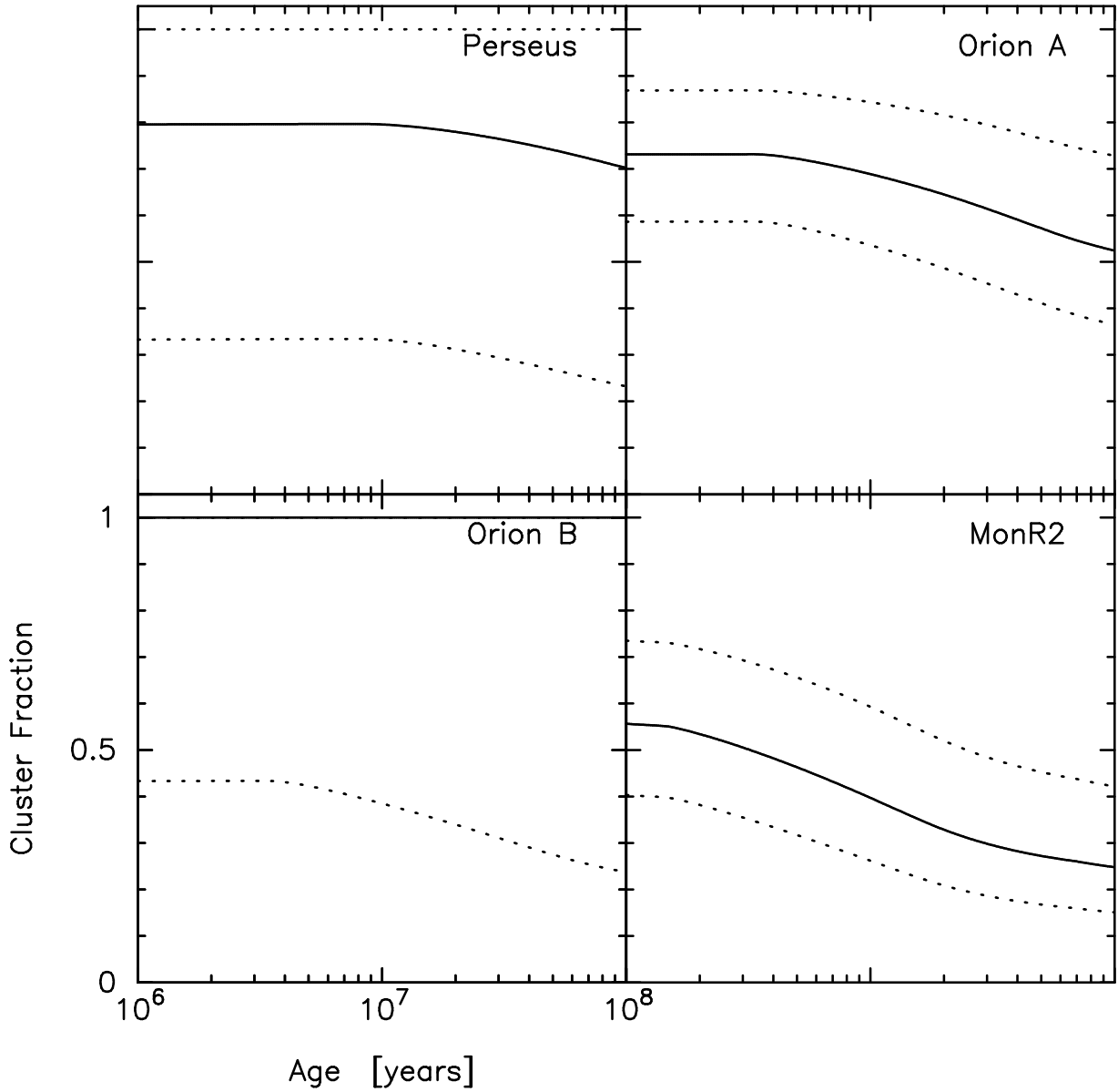

\insertplot{figure21.ps}{5.0}{4}{-1.5}{3.0}{0.9}{1}
\vspace{1.2truein}
\caption{
   The fraction of the total stellar population currently contained in 
   clusters as a function of the cloud age assuming that only part of the 
   distributed population is detected at a given age as indicated by the 
   model calculations shown in Figure~\ref{fraction}. The solid lines show 
   the fraction of stars in clusters for the nominal extinction model, and the 
   dotted lines represent the low and high extinction models. 
  \label{clusterfraction}
}
\end{figure}
\clearpage




\tablenum{1}

\begin{deluxetable}{cccccccc@{\extracolsep{10pt}}cc}
\tablewidth{470pt}
\tablecaption{Molecular Cloud Properties\label{tbl:clouds}}
\tablehead{
\colhead{Cloud}          & 
\multicolumn{2}{c}{Coordinates} &
\colhead{Distance}       & 
\colhead{Area}           & 
\multicolumn{3}{c}{Mass} &
\multicolumn{2}{c}{References}\\
\cline{2-3}
\cline{6-8}
\cline{9-10}\\
\\[-8.0ex]\\
\colhead{}         &
\colhead{$\ell$}   &
\colhead{$b$}      &
\colhead{}         &
\colhead{}         &
\colhead{Low}      &
\colhead{Nominal}  &
\colhead{High}     &
\colhead{Distance} &
\colhead{\thco}\\
\colhead{}         &
\colhead{}         &
\colhead{}         &
\colhead{[pc]}     &
\colhead{[deg$^2$]}  &
\multicolumn{3}{c}{[10$^4$ M$_\odot$]}
}
\startdata
Perseus & 160 & $-19$ & 320 & 6.4 & 1.0 & 1.3 & 1.9 & 1 & 4\\
Orion~A & 211 & $-20$ & 480 & 6.2 & 2.5 & 3.3 & 4.8 & 2 & 5\\
Orion~B & 206 & $-15$ & 480 & 7.2 & 2.4 & 3.2 & 4.7 & 2 & 6\\
MonR2   & 214 & $-13$ & 830 & 2.6 & 2.3 & 3.1 & 4.6 & 3 & 6\\
\enddata
\tablerefs{\\
  (1) de Zeeuw \etal (1999)\\
  (2) Genzel \etal (1981)\\
  (3) Herbst \& Racine (1976)\\
  (4) Padoan \etal (1999)\\
  (5) Bally \etal (1987)\\
  (6) Miesch \& Bally (1994)\\
}
\end{deluxetable}



\newcommand{\nd}{\multicolumn{1}{c}{\nodata}}

\tablenum{2}
\pagestyle{empty}

\begin{deluxetable}{ccc@{\extracolsep{20pt}}c@{\extracolsep{10pt}}c@{\extracolsep{10pt}}rcl}
\tablewidth{490pt}
\tablecaption{Stellar Clusters\label{tbl:clusters}}
\tablehead{
\colhead{ID} & \multicolumn{2}{c}{Galactic} & 
\multicolumn{2}{c}{Equatorial (J2000)} &
\multicolumn{1}{c}{N$_{\rm stars}$} & \colhead{R$_{\rm eff}$} & 
\multicolumn{1}{c}{Comments}\\
\cline{2-3}
\cline{4-5}\\
\colhead{}         &
\colhead{$\ell$}   &
\colhead{$b$}      &
\colhead{$\alpha$} &
\colhead{$\delta$} &
\colhead{}         &
\colhead{[pc]}     &
}
\startdata
\multicolumn{8}{c}{\it Perseus}\\
 1 & 160.5044 & -17.8011  & 03:44:37.2 & +32:09:19 & $>$299 & 1.19 & IC 348\\
\\
\multicolumn{8}{c}{\it Orion A}\\
 2 & 208.9740 & $-$19.3458 & 05:35:20.9 & $-$05:20:29 & $>$1744 & 3.86 & Orion Nebula Cluster \\
 3 & 210.8003 & $-$19.5051 & 05:37:52.9 & $-$06:57:09 &   23 & 0.56 & V592 Ori\\
 4 & 210.9800 & $-$19.3276 & 05:38:49.6 & $-$07:01:32 &   36 & 0.61 & L1641 C\\
 5 & 212.4710 & $-$19.0181 & 05:42:27.1 & $-$08:09:12 &  129 & 1.10 & L1641 South\\
 6 & 212.9816 & $-$19.1469 & 05:42:50.1 & $-$08:38:37 &   15 & 0.42 & IRAS 05404-0839\\
 7 & 214.0556 & $-$19.6177 & 05:42:53.5 & $-$09:45:38 &   23 & 0.47 & IRAS 05404-0946\\
 8 & 214.2706 & $-$19.7824 & 05:42:38.4 & $-$10:00:51 &   22 & 0.47 & IRAS 05401-1002\\
\\
\multicolumn{8}{c}{\it Orion B}\\
 9 & 205.3080 & $-$14.2957 & 05:46:47.0 & +00:06:16 & $>$45 & 0.72 & NGC 2068\\
10 & 206.5122 & $-$16.3719 & 05:41:37.3 & $-$01:53:39 & $>$201 & 1.01 & NGC 2024 \\
\\
\multicolumn{8}{c}{\it MonR2}\\
11 & 213.3381 & $-$12.6029  & 06:07:08.1 & $-$06:03:53 &  15 & 0.41 & IRAS 06046-0603\\
12 & 213.6955 & $-$12.5926  & 06:07:47.8 & $-$06:22:20 & 371 & 1.85 & MonR2\\
13 & 213.8745 & $-$11.8410  & 06:10:49.1 & $-$06:11:38 & 134 & 1.13 & GGD 12-15\\
14 & 214.1337 & $-$11.4173  & 06:12:48.0 & $-$06:13:56 &  23 & 0.61 & GGD 17\\
\enddata
\end{deluxetable}




\tablenum{3}
\pagestyle{empty}

\begin{deluxetable}{c@{\extracolsep{10pt}}ccc@{\extracolsep{10pt}}ccc}
\tablewidth{380pt}
\tablecaption{The Distributed Stellar Population\label{tbl:distributed}}
\tablehead{
\colhead{Cloud}   & 
\multicolumn{3}{c}{Surface Density (arcmin$^{-2}$)} & 
\multicolumn{3}{c}{N$_{\rm stars}$} \\
\cline{2-4}
\cline{5-7}\\
\colhead{}        & 
\colhead{Low}     &
\colhead{Nominal} &
\colhead{High}    &
\colhead{Low}     &
\colhead{Nominal} &
\colhead{High}
}
\startdata
Perseus\\
\multicolumn{1}{c}{$J$}   & -0.048 & 0.013 & 0.101 & \nd & 310 & 2300\\
\multicolumn{1}{c}{$H$}   & -0.031 & 0.006 & 0.064 & \nd & 150 & 1500\\
\multicolumn{1}{c}{$K_s$} & -0.014 & 0.004 & 0.034 & \nd & 100 &  780\\
\\
Orion A\\
\multicolumn{1}{c}{$J$}   & -0.054 & 0.009 & 0.099 & \nd & 210 & 2200\\
\multicolumn{1}{c}{$H$}   & -0.017 & 0.021 & 0.080 & \nd & 470 & 1800\\
\multicolumn{1}{c}{$K_s$} &  0.013 & 0.032 & 0.064 & 300 & 730 & 1400\\
\\
Orion B\\
\multicolumn{1}{c}{$J$}   & -0.10  & -0.018 & 0.096 & \nd  & \nd & 2500\\
\multicolumn{1}{c}{$H$}   & -0.057 & -0.008 & 0.066 & \nd  & \nd & 1700\\
\multicolumn{1}{c}{$K_s$} & -0.033 & -0.009 & 0.031 & \nd  & \nd &  790\\
\\
MonR2\\
\multicolumn{1}{c}{$J$}   & -0.031 &  0.070 & 0.205 & \nd  &  660 & 1900\\
\multicolumn{1}{c}{$H$}   &  0.001 &  0.056 & 0.136 &   10 &  530 & 1300\\
\multicolumn{1}{c}{$K_s$} &  0.020 &  0.045 & 0.083 &  190 &  420 &  780\\
\enddata
\end{deluxetable}

\end{document}